\documentclass[letterpaper,twocolumn,showpacs,preprintnumbers,amsmath,amssymb,floatfix,superscriptaddress]{revtex4}

\usepackage{graphicx}  % Include figure files
\usepackage{dcolumn}   % Align table columns on decimal point
\usepackage{bm}        % bold math
\usepackage{epsfig}

\begin{document}

\title{Variational calculation of the limit cycle and its frequency \\in a two-neuron model with delay}

\author{Sebastian F.~Brandt}
\email{sbrandt@physics.wustl.edu}
\affiliation{
Department of Physics, Campus Box 1105, Washington University in St.~Louis, MO 63130-4899, USA}

\author{Axel Pelster}
\email{axel.pelster@uni-duisburg-essen.de}
\affiliation{%                                    
Universit{\"a}t Duisburg-Essen, Campus Essen, Fachbereich Physik, Universit{\"a}tsstra{\ss}e 5, 45117 
Essen, Germany}%      

\author{Ralf Wessel}
\affiliation{
Department of Physics, Campus Box 1105, Washington University in St.~Louis, MO 63130-4899, USA}

\date{April 25, 2006}
\begin{abstract}
We consider a model system of two coupled Hopfield neurons, which is described by delay differential equations taking into account the finite signal propagation and processing times. When the delay exceeds a critical value, a limit cycle emerges via a supercritical Hopf bifurcation. First, we calculate its frequency and trajectory perturbatively by applying the Poincar{\'e}-Lindstedt method. Then, the perturbation series are resummed by means of the Shohat expansion in good agreement with numerical values. However, with increasing delay, the accuracy of the results from the Shohat expansion worsens. We thus apply variational perturbation theory (VPT) to the perturbation expansions to obtain more accurate results, which moreover hold even in the limit of large delays. 
\end{abstract}
\pacs{82.40.Bj, 84.35.+i, 02.30.Ks, 02.30.Mv}
\maketitle
%
%%%%%%%%%%%%%%%%%%%%%%%%%%%%%%%%%%%%%%%%
\section{Introduction}
%%%%%%%%%%%%%%%%%%%%%%%%%%%%%%%%%%%%%%%%
Feedback in biological systems has received increased attention in recent years \cite{Bechhoefer}.  In particular, the role of delayed recurrent loops in models of population dynamics, epidemiology, physiology, immunology, neural networks, and cell kinetics has been studied extensively \cite{Bocharov}. Neural network systems are complex and large-scale nonlinear dynamical systems, and the dynamics of a delayed network are yet richer and more complicated \cite{Wu}.  Hopfield \cite{Hopfield} proposed a simplified model of a neural network in which each neuron is represented by a linear circuit consisting of a resistor and a capacitor, coupled to other neurons via nonlinear sigmoidal activation functions.  From this model, he derived a system of first-order ordinary differential equations to describe the network dynamics.  Extending Hopfield's model, Marcus and Westervelt \cite{Marcus} considered the effect of including a temporal delay in the model to account for finite propagation and signal processing times. 

In networks of real neurons, delays occur at the synaptic level due to transmitter release dynamics and the integration time of post-synaptic potentials at the dendritic tree level where post-synaptic potentials have a finite conduction speed to the soma, and in the axons due to the finite axonal conduction speed of action potentials \cite{Eurich}.  It is well-known that time delay can cause an otherwise stable system to oscillate \cite{Heiden,Coleman,Hadeler} and may lead to bifurcation scenarios resulting in chaotic dynamics \cite{Wischert,Schanz}. 
On the other hand, delayed feedback permits the control of chaos \cite{OGY}, where it can be used to stabilize unstable periodic orbits in chaotic attractors \cite{Pyragas1,Pyragas2}. Experimentally, time-delayed chaos control was successfully applied, for instance,
to electronic oscillators \cite{Pyragas3}, mechanical pendula \cite{Christini}, lasers \cite{Bielawski}, and chemical systems \nolinebreak[4]\cite{Parmananda}. Furthermore, a recently proposed scheme for the treatment of neurological disorders employs delayed feedback in order to efficiently desynchronize the activity of oscillatory neurons \cite{Tass}. Therefore, finite delays are an essential property of any realistic model of a neuron population \cite{Milton}.

In the vast majority of cases, information about a physical system can only be obtained by
means of numerical or analytical approximation methods.  Numerical methods constitute a powerful and effective tool to describe even extremely complicated physical scenarios.  Nevertheless, their accuracy is not always superior to that of analytical approximations, and usually more insight into the physical principles that govern the system is obtained by pursuing an analytical approach. Often, perturbation expansions are easily accessible, but they are usually divergent and need resummation. 
A recently developed, powerful method to perform such a resummation is variational perturbation theory (VPT), which has been successfully applied in various quantum or statistical field theories \cite{Feynman2,Kleinertsys,PathInt3,VerenaBuch,Festschrift}. 
A first application of VPT in the field of deterministic nonlinear dynamics is found in Ref.~\cite{Schanz2}, while the present work extends the use of VPT for the first time to a system described by delay differential equations (DDE's).
\begin{figure*}[t]
  \begin{center}
    \epsfig{file=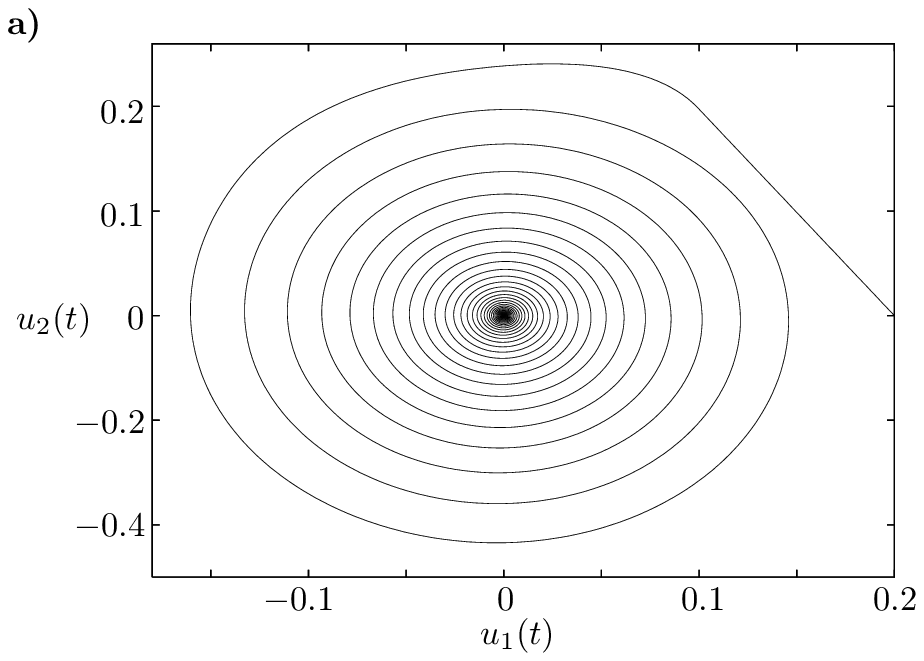,width=\columnwidth} 
    \hfill
    \epsfig{file=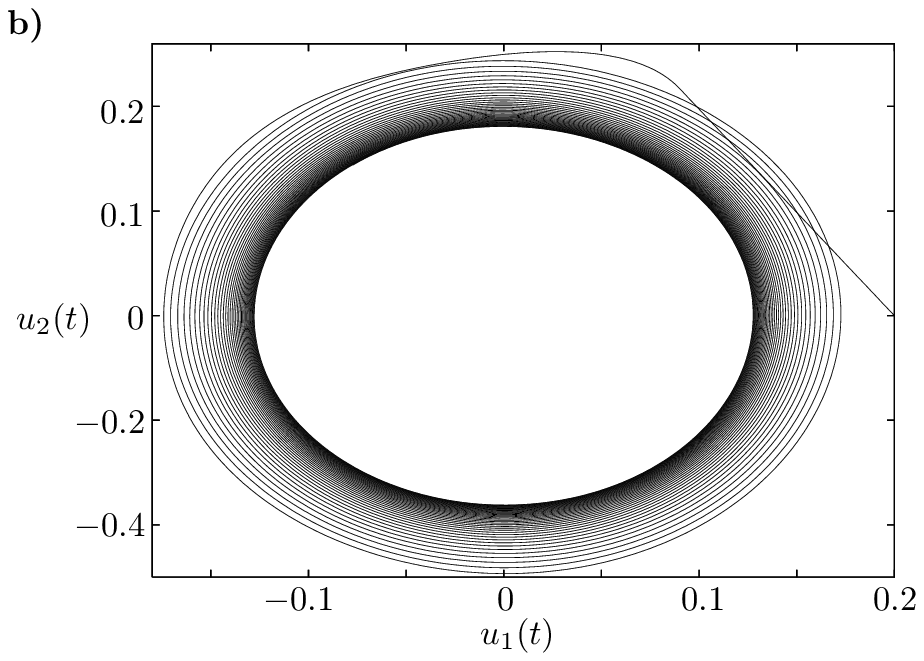,width=\columnwidth}
    \caption{Numerical solutions of the system of DDE's (\ref{dde1}), (\ref{dde2}) with $a_1 = -1$, $a_2 = 2$ and $\tau^{(1)} = \tau^{(2)} = \tau$. For this choice of parameters, the critical value of the delay is $\tau_0 = \pi / 4 \approx 0.7854 \ldots \, .$ In {\bf a)} the delay is $\tau = 0.7$, and the origin is a stable fixed point.  In {\bf b)} the delay exceeds the critical value:  $\tau = 0.8$. In this case, the origin is unstable and the trajectory approaches a limit cycle.  In both cases the initial conditions are  $u_1(t) = 0.2$, $u_2(t) = 0$ for $t\in[-\tau, 0]$.}
    \label{fig1}
  \end{center}
\end{figure*}   

In Sec.~\ref{Model}, we introduce the two-neuron model and the system of DDE's that we consider. The results of a linear stability analysis of the model system are reported in Sec.~\ref{LSA}, and it is shown that a limit cycle emerges via a supercritical Hopf bifurcation when the delay exceeds a critical value. In Sec.~\ref{PLM}, the Poincar{\'e}-Lindstedt Method is applied to derive the perturbation expansions for the delay-induced limit cycle and its angular frequency. In Sec.~\ref{SE}, we apply the Shohat expansion to the perturbation series of the limit cycle and its angular frequency as a first crude resummation approach. In Sec.~\ref{VPT}, we resum the perturbation expansions using VPT, which allows us to improve the quality of our results significantly and to obtain results which are reasonable even in the limit of large delays.
%%%%%%%%%%%%%%%%%%%%%%%%%%%%%%%%%%%%%%%%%%%%%%%%%%%%%%
\section{Model} \label{Model}
%%%%%%%%%%%%%%%%%%%%%%%%%%%%%%%%%%%%%%%%%%%%%%%%%%%%%%
Neural circuits composed of two or three neurons form the basic feedback mechanisms involved in the regulation of neural activity \cite{Milton}. Many researchers have used bifurcation analysis and numerical simulations in order to analyze a system of two Hopfield-like neurons with discrete or distributed time delays \cite{Babcock,Wei,Ruan,Shayer,Gopalsamy,Liao,Olien,Majee,Liao2}. In this investigation, we apply analytical approximation methods to a two-neuron system with delay, described by the coupled first-order DDE's
\begin{eqnarray}
\frac{d u_1 (t)}{dt} &=& -u_1(t) + a_1 \tanh[u_2(t - \tau^{(2)})] \label{dde1} \\
\frac{d u_2 (t)}{dt} &=& -u_2(t) + a_2 \tanh[u_1(t - \tau^{(1)})] \label{dde2} \,.
\end{eqnarray}
Here,  $u_1$ and $u_2$ denote the voltages of the Hopfield neurons and $\tau^{(1)}$ and $\tau^{(2)}$ are the signal propagation or processing time delays, while $a_1$ and $a_2$ describe the couplings between the two neurons. 
%%%%%%%%%%%%%%%%%%%%%%%%%%%%%%%%%%%%%%%%%%%%%%%%%%%%%%%%%
\section{Linear Stability Analysis} \label{LSA}
%%%%%%%%%%%%%%%%%%%%%%%%%%%%%%%%%%%%%%%%%%%%%%%%%%%%%%%%%
The system of DDE's (\ref{dde1}), (\ref{dde2}) has a trivial stationary point at $u_1 = u_2 = 0$ and we first analyze its stability.  Near the equilibrium point, linearizing the DDE system yields
\begin{eqnarray}
\dot{u}_1(t) &=& -u_1(t) + a_1 u_2(t - \tau^{(2)}) \,, \\
\dot{u}_2(t) &=& -u_2(t) + a_2 u_1(t - \tau^{(1)}) \,.
\end{eqnarray}
Setting
\begin{eqnarray}
\mathbf{u}(t) = e^{\lambda t}
\left(
  \begin{array}{ c }
     c_1 \\
     c_2
  \end{array} 
\right)
\end{eqnarray}
in the last equation, where $\lambda$ is a complex number, and $c_1$ and $c_2$ are constants, we get a nontrivial solution if and only if
\begin{eqnarray}
(\lambda + 1)^2 - a_1 a_2 e^{-\lambda (\tau^{(1)} + \tau^{(2)})} = 0\,. \label{eqc}
\end{eqnarray}
This equation has been analyzed in detail in Ref.~\cite{Wei}.  For $a_1 a_2 \leq -2$ the conditions of Theorem 2 in Ref.~\cite{Wei} are met. Defining $\tau = (\tau^{(1)} + \tau^{(2)})/2$ and
\begin{eqnarray}
\tau_{j} \equiv \frac{1}{2 \omega_0}\left[ \sin^{-1}\left(-\frac{2 \omega_0}{a_1a_2}\right) + 2 j \pi \right]\hspace{-0.5mm}, \quad \hspace{-2mm} j = 0,\,1,\,2,\, \ldots ,  
\end{eqnarray}
where $\omega_0 = \sqrt{|a_1 a_2| - 1}$, this theorem states that:
\begin{itemize}
\item If $\tau \in [0,\, \tau_0)$, then the zero solution of (\ref{dde1}), (\ref{dde2}) is asymptotically stable.
\item If $\tau > \tau_0$, then the zero solution of (\ref{dde1}), (\ref{dde2}) is unstable.
\item $\tau_j$, with $j = 0,\, 1,\, 2\, \ldots$, are Hopf bifurcation values of (\ref{dde1}), (\ref{dde2}).
\end{itemize}
Furthermore, Theorem 3 in Ref.~\cite{Wei} states that the Hopf bifurcation at $\tau = \tau_0$ is supercritical.
Note that $i \omega_0$ is the solution to (\ref{eqc}) when $\tau = \tau_0$, and the period of the limit cycle at the Hopf bifurcation is thus $T_0 = 2 \pi / \omega_0$.

%%%%%%%%%%%%%%%%%%%%%%%%%%%%%%%%%%%%%%%%%%
\section{Poincar{\'e}-Lindstedt Method} \label{PLM}
%%%%%%%%%%%%%%%%%%%%%%%%%%%%%%%%%%%%%%%%%%%
Figure \ref{fig1} shows numerical solutions of the system of DDE's (\ref{dde1}), (\ref{dde2}) for the two cases in which the delay $\tau$ is either smaller or greater than its critical value.
Below the critical value $\tau_0$ of the delay $\tau$ no periodic solution exists, while above $\tau = \tau_0$ there is such a solution.  We now consider the case $\tau^{(1)} = \tau^{(2)} = \tau$, $a_1 a_2 \leq -2$ and seek to calculate the period and trajectory of the periodic solution approximatively.  To this end, we apply the Poincar{\'e}-Lindstedt method \cite{MacDonald}.  Since a supercritical Hopf bifurcation occurs at $\tau = \tau_0$, we assume that the amplitude and frequency of the new periodic states are analytic in $\epsilon = \sqrt{\tau - \tau_0}$ and expand them as
\begin{eqnarray}
\mathbf{u}(t) &=& \epsilon \mathbf{U}(t) = \epsilon \left[ \mathbf{U}^{(0)}(t) + \epsilon \mathbf{U}^{(1)}(t) + \ldots \right] \,, \label{uexp} \\
\omega(\epsilon) &=& \omega_0 + \epsilon \omega_1 + \epsilon^2 \omega_2 + \ldots \, .
\end{eqnarray}
It is convenient to rescale the argument of these functions so that they become periodic with period $2\pi$.  We thus introduce the new independent variable $\xi$ according to
\begin{eqnarray}
\xi = \omega(\epsilon) t \, , \label{tscal}
\end{eqnarray}
and we write
\begin{eqnarray}
\mathbf{U}(t) = \mathbf{V}(\xi)\, . \label{UV}
\end{eqnarray}
Applying the perturbation expansion (\ref{uexp}) to the system of DDE's (\ref{dde1}), (\ref{dde2}) and performing the change of variables (\ref{tscal}), (\ref{UV}), we obtain
%%%%%%%%%%%%%%%%%%%%%%%%%%%%%%%%%%%%%%%%%%%%%%%%%%%%%%%%%%%%%%%%%%%%%%%%%%%%%
\begin{table*}[t]
\begin{center}
\begin{tabular}{|r|c|}
\hline
\rule[-1mm]{0mm}{4mm}\hspace*{1mm}$n$\hspace*{1mm}& $\omega_n$ \\[1mm] \hline
\rule[-1mm]{0mm}{7mm}\hspace*{1mm}$2$\hspace*{1mm} & ${\displaystyle -\frac{4}{2 + \pi} }$
\\[3mm] \hline
\rule[-1mm]{0mm}{7mm}\hspace*{1mm}$4$\hspace*{1mm}& ${\displaystyle \frac{4(341 + 108 \pi)}{27(2 + \pi)^3} }$ \\[3mm] \hline
\rule[-1mm]{0mm}{7mm}\hspace*{1mm}$6$\hspace*{1mm}& $ {\displaystyle -\frac{8(73843 + 40773 \pi + 5832 \pi^2)}{729(2 + \pi)^5} }$ \\ [3mm]\hline
\rule[-1mm]{0mm}{7mm}\hspace*{1mm}$8$\hspace*{1mm} & $ \hspace*{1mm}{\displaystyle \frac{1440729464 + 3 \pi(359606308 + 92814567 \pi + 8398080 \pi^2)}{98415(2 + \pi)^7} }$ \hspace*{1mm}\\[3mm] \hline
\rule[-1mm]{0mm}{7mm}\hspace*{1mm}$10$\hspace*{1mm} & $ \hspace*{1mm}{\displaystyle - \frac{2 (1885638326848 + 9 \pi (193375795408 + 3 \pi (22966214893 + 4 \pi (952738307 + 62985600 \pi))))}{13286025(2 + \pi)^9} }$ \hspace*{1mm}\\[3mm] \hline
\rule[-1mm]{0mm}{5mm}\hspace*{1mm}$12$\hspace*{1mm} & $ \hspace*{1mm}{\displaystyle ( 48294520193761504 + 3 \pi (17432699637100336 + 3 \pi (2577825095210584} $ \\ & $ \hspace*{1mm}{\displaystyle + \pi(596088219927028 + 72959354094441 \pi + 3809369088000 \pi^2))))/(8370195750(2 + \pi)^{11}) }$ \hspace*{1mm}\\[1mm] \hline
\rule[-1mm]{0mm}{5mm}\hspace*{1mm}$14$\hspace*{1mm} & $ \hspace*{1mm}{\displaystyle -(137083613818976067424 + 
    3 \pi (56352224911533618320 + 
          3 \pi (9835626348748269040 + 3 \pi (949130678879606440 } $ \\ & $ \hspace*{1mm}{\displaystyle +3 \pi (54285350368574420 + \pi (5287281140608997 + 228562145280000 \pi)))))) /(1129976426250 (2 + \pi)^{13}) }$ \hspace*{1mm}\\[1mm] \hline
\rule[-1mm]{0mm}{5mm}\hspace*{1mm}$16$\hspace*{1mm} & $ \hspace*{1mm}{\displaystyle  (290578164278923471719089408 + 
    9 \pi (44452665928743252091582336 + 
          3 \pi (8868376426577693217600640 } $ \\ & $ \hspace*{1mm}{\displaystyle + 3 \pi (1013305929108995195501920 +  9 \pi (24272564564656648301080 + \pi 
(3331148075811324207916 }$ \\ & $ \hspace*{1mm}{\displaystyle + \pi (270489187825118497343 + 9983594505830400000 \pi ))))))) / (111054083171850000 (2+\pi )^{15})}$ \hspace*{1mm}\\[1mm] \hline\end{tabular}
\caption{Expansion coefficients for the angular frequency of the limit cycle for $a_1 = -1$, $a_2 = 2$ up to order $\epsilon^{16}$.}
\label{OmTab}
\end{center}
\end{table*}
%%%%%%%%%%%%%%%%%%%%%%%%%%%%%%%%%%%%%%%%%%%%%%%%%%%%%%%%%%
\begin{eqnarray}
\hspace{-4mm} \omega(\epsilon) \frac{d V_{1}(\xi)}{d \xi} &=&
 - V_{1}(\xi) + \frac{a_{1}}{\epsilon} \tanh \left\{ \epsilon V_{2}[\xi - \alpha(\epsilon)] \right \} \, , \label{ddesys1} \\
\hspace{-4mm} \omega(\epsilon) \frac{d V_{2}(\xi)}{d \xi} &=&
 - V_{2}(\xi) + \frac{a_{2}}{\epsilon} \tanh \left\{ \epsilon V_{1}[\xi - \alpha(\epsilon)] \right \} \, ,  \label{ddesys2}
\end{eqnarray}
in which
\begin{eqnarray}
\alpha(\epsilon) &=& \omega(\epsilon) \tau = \omega(\epsilon)(\tau_0 + \epsilon^2) \nonumber \\
&=& \omega_0 \tau_0 + \epsilon \omega_1 \tau_0  + \epsilon^2(\omega_0 +  \omega_2 \tau_0) + \ldots \, . \label{alphaexp}
\end{eqnarray}
The delayed variable $V_{1/2}[\xi - \alpha(\epsilon)]$ is written as
\begin{eqnarray}
\mathbf{V}[\xi - \alpha(\epsilon)] = \mathbf{V}^{(0)}(\xi, \alpha) + \epsilon \mathbf{V}^{(1)}(\xi, \alpha) + \ldots \, ,
\end{eqnarray}
corresponding to the expansion in (\ref{uexp}), which is equivalent to
\begin{eqnarray}
\mathbf{V}(\xi) = \mathbf{V}^{(0)}(\xi) + \epsilon \mathbf{V}^{(1)}(\xi) + \ldots \, .
\end{eqnarray}
In order to take into account (\ref{alphaexp}), each term in the expansion of $\mathbf{V}(\xi - \alpha)$ is expanded as a Taylor series:
\begin{eqnarray}
\mathbf{V}^{(j)}(\xi, \alpha) &=& \\ && \hspace{-12mm} \mathbf{V}^{(j)}(\xi - \omega_0 \tau_0) - \epsilon  \omega_1 \tau_0 \left. \frac{d \mathbf{V}^{(j)}(\xi')}{d \xi'} \right|_{\xi' = \xi - \omega_0 \tau_0 } + \ldots \,. \nonumber
\end{eqnarray}
Applying the expansions for $\mathbf{V}(\xi)$ and $\mathbf{V}(\xi - \alpha)$ to (\ref{ddesys1}), (\ref{ddesys2}), we obtain to zeroth order in $\epsilon$
\begin{eqnarray}
\frac{d V_1^{(0)}(\xi)}{d \xi} &=& - \frac{V_1^{(0)}(\xi)}{\omega_0} + \frac{a_1}{\omega_0}V_2^{(0)}(\xi - \omega_0 \tau_0) \, , \label{V10DE} 
\end{eqnarray}
\pagebreak[4]
\begin{eqnarray}
\frac{d V_2^{(0)}(\xi)}{d \xi} &=& - \frac{V_2^{(0)}(\xi)}{\omega_0} + \frac{a_2}{\omega_0} V_1^{(0)}(\xi - \omega_0\tau_0) \, . \label{V20DE}
\end{eqnarray}
Imposing the initial conditions $V_1^{(0)}(0) = A_0$, $V_2^{(0)}(0) = B_0$ on the periodic solution $\mathbf{V}^{(0)}(\xi)$, we find the general solution to the system of homogeneous differential equations (\ref{V10DE}), (\ref{V20DE}) as
\begin{eqnarray}
V_1^{(0)}(\xi) &=& A_0 \cos \xi + B_0 a_1 \sin (\omega_0 \tau_0 ) \sin \xi \, , \label{V10} \\ 
V_2^{(0)}(\xi) &=&  B_0 \cos \xi - \frac{A_0}{a_1 \sin(\omega_0 \tau_0 )} \sin \xi \, .  \label{V20}
\end{eqnarray}
The periodic solution $\mathbf{V}(\xi)$ to (\ref{ddesys1}), (\ref{ddesys2}) can only be determined up to an arbitrary phase.  Without loss of generality we can thus choose $B_0 = 0$ in (\ref{V10}), (\ref{V20}), which fixes the phase of the zeroth order solution, at least up to a shift of $\pi$.

In general, to order $\epsilon^n$, we have to solve the system of differential equations 
\begin{eqnarray}
\hspace{-5mm} \frac{d V_1^{(n)}(\xi)}{d \xi} &=& - \frac{V_1^{(n)}(\xi)}{\omega_0} +\frac{a_1}{\omega_0} V_2^{(n)}(\xi - \omega_0 \tau_0) + f_1^{(n)}(\xi)\, , \nonumber \hspace{-4mm} \\ \label{V1nDE} \\
\hspace{-5mm} \frac{d V_2^{(n)}(\xi)}{d \xi} &=& - \frac{V_2^{(n)}(\xi)}{\omega_0} + \frac{a_2}{\omega_0} V_1^{(n)}(\xi - \omega_0 \tau_0) + f_2^{(n)}(\xi)\, ,  \nonumber \hspace{-4mm} \\ \label{V2nDE}
\end{eqnarray}
where the inhomogeneity $\mathbf{f}^{(n)}(\xi)$ is determined by the solutions to previous orders.  Since we require that the solution $\mathbf{V}^{(n)}(\xi)$ be periodic in $\xi$ with period $2\pi$, we can impose certain conditions on the inhomogeneity $\mathbf{f}^{(n)}(\xi)$.  Namely, we demand that $\mathbf{f}^{(n)}(\xi)$ not contain terms that would lead to non-periodic solutions for $\mathbf{V}^{(n)}(\xi)$, i.e., $\mathbf{f}^{(n)}(\xi)$ must not contain secular terms.  In order to identify the conditions that must be satisfied by $\mathbf{f}^{(n)}(\xi)$, we expand $\mathbf{V}^{(n)}(\xi)$ and $\mathbf{f}^{(n)}(\xi)$ as Fourier series:
\begin{eqnarray}
\left(
  \begin{array}{ c }
     V_1^{(n)}(\xi) \\
     V_2^{(n)}(\xi)
  \end{array}
\right)
& \hspace{-2mm}=& \hspace{-1mm}\sum_{k = 1}^{\infty}\left[
\left(
   \begin{array}{ c }
     a_{1, k}^{(n)} \\
     a_{2, k}^{(n)}
  \end{array}
\right)
\cos k \xi
+
\left(  
\begin{array}{ c }
     b_{1, k}^{(n)} \\
     b_{2, k}^{(n)}
\end{array}
\right) 
\sin k \xi
\right]\, , \nonumber \\ \label{VFouExp}\\
\left(
  \begin{array}{ c }
     f_1^{(n)}(\xi) \\
     f_2^{(n)}(\xi)
  \end{array}
\right)
& \hspace{-2mm}=& \hspace{-1mm} \sum_{k = 1}^{\infty}\left[
\left(
   \begin{array}{ c }
     \alpha_{1, k}^{(n)} \\
     \alpha_{2, k}^{(n)}
  \end{array}
\right)
\cos k \xi
+
\left(  
\begin{array}{ c }
     \beta_{1, k}^{(n)} \\
     \beta_{2, k}^{(n)}
\end{array}
\right)
\sin k \xi
\right] \, . \nonumber \label{fFouExp} \\
\end{eqnarray}
By inserting the expansions (\ref{VFouExp}), (\ref{fFouExp}) into the systems of equations (\ref{V1nDE}), (\ref{V2nDE}), we find that the coefficient of the the terms with $k=1$ in the inhomogeneity ${\bf f}^{(n)}(\xi)$ must satisfy the conditions
\begin{eqnarray}
a_2 \sin(\omega_0 \tau_0) \alpha_{1,1}^{(n)} + \beta_{2,1}^{(n)} &=& 0 \, , \label{Cond1} \\
\alpha_{2,1}^{(n)} - a_2 \sin(\omega_0 \tau_0) \beta_{1,1}^{(n)} &=& 0 \, .  \label{Cond2} 
\end{eqnarray}
The derivation of these two conditions is demonstrated in the appendix.

After this general result, we now consider the first-order expansion of the system (\ref{ddesys1}), (\ref{ddesys2}). Taking into account the result (\ref{V10}), (\ref{V20}) and the choice $B_0 = 0$, we obtain the inhomogeneity ${\mathbf f}^{(1)}$ to be given by
\begin{eqnarray}
f_1^{(1)}(\xi) = A_0 \omega_1 \left( \tau_0 \cos \xi + \frac{1 + \tau_0}{\omega_0}\sin \xi \right)
\end{eqnarray}
and
%%%%%%%%%%%%%%%%%%%%%%%%%%%%%%%%%%%%%%%%%%%%%%%%%%%%%%%%%%
\begin{table*}[t]
\begin{center}
\begin{tabular}{|c|c|c||c|c|c|}
\hline
\rule[-1mm]{0mm}{7mm} \hspace*{1mm} $a_{1,k}^{(n)}$ \hspace*{1mm} & \hspace*{1mm} $k=1$\hspace*{1mm} & \hspace*{1mm} $k=3$ \hspace*{1mm} & \hspace*{1mm} $b_{1,k}^{(n)}$\hspace*{1mm} & \hspace*{1mm} $k = 1$ \hspace*{1mm} & \hspace*{1mm} $k = 3$ \hspace*{1mm} \\[3mm] \hline
\rule[-1mm]{0mm}{7mm} \hspace*{1mm} $n = 0$ \hspace*{1mm} & \hspace*{1mm} ${\displaystyle \frac{4}{\sqrt{3(2 + \pi)} } }$ \hspace*{1mm} & \hspace*{1mm} $0$ \hspace*{1mm} & \hspace*{1mm} $n = 0$ \hspace*{1mm} & \hspace*{1mm} $0$ \hspace*{1mm} & \hspace*{1mm} $0$
\hspace*{1mm} \\[3mm] \hline
\rule[-1mm]{0mm}{7mm} \hspace*{1mm} $n=2$ \hspace*{1mm} & \hspace*{1mm} ${\displaystyle -\frac{5\sqrt{3}(116 + 33 \pi)}{81(2 + \pi)^{5/2} } }$ 
\hspace*{1mm} & \hspace*{1mm} ${\displaystyle - \frac{2\sqrt{3}}{27(2 + \pi)^{3/2} } }$ 
\hspace*{1mm} & \hspace*{1mm} $n = 2$ \hspace*{1mm} & \hspace*{1mm} 0 \hspace*{1mm} & \hspace*{1mm} ${\displaystyle \frac{14 \sqrt{3} }{27(2 + \pi)^{3/2}} }$ \hspace*{1mm} \\[3mm] \hline \hline
\rule[-1mm]{0mm}{7mm} \hspace*{1mm} $a_{2,k}^{(n)}$ \hspace*{1mm} & \hspace*{1mm} $k=1$\hspace*{1mm} & \hspace*{1mm} $k=3$ \hspace*{1mm} & \hspace*{1mm} $b_{2,k}^{(n)}$\hspace*{1mm} & \hspace*{1mm} $k = 1$ \hspace*{1mm} & \hspace*{1mm} $k = 3$ \hspace*{1mm} \\[3mm] \hline
\rule[-1mm]{0mm}{7mm} \hspace*{1mm} $n = 0$ \hspace*{1mm} & \hspace*{1mm} $0$ \hspace*{1mm} & \hspace*{1mm} $0$ \hspace*{1mm} & \hspace*{1mm} $n = 0$ \hspace*{1mm} & \hspace*{1mm} ${\displaystyle \frac{4 \sqrt{2}}{\sqrt{3(2 + \pi)} } }$ \hspace*{1mm} & \hspace*{1mm} $0$
\hspace*{1mm} \\[3mm] 
\hline
\rule[-1mm]{0mm}{7mm} \hspace*{1mm} $n=2$ \hspace*{1mm} & \hspace*{1mm} $0$ \hspace*{1mm} & \hspace*{1mm} ${\displaystyle  \frac{10\sqrt{6}} {27(2 + \pi)^{3/2}} }$ 
\hspace*{1mm} & \hspace*{1mm} $n = 2$ \hspace*{1mm} & \hspace*{1mm} ${\displaystyle -\frac{\sqrt{6}(436 + 93 \pi)}{81(2 + \pi)^{5/2}}}$ \hspace*{1mm} & \hspace*{1mm}
${\displaystyle -\frac{2\sqrt{6}}{27(2 + \pi)^{3/2}} }$ \hspace*{1mm} \\[3mm] \hline
\end{tabular}
\caption{Fourier expansion coefficients of the limit cycle for $a_1 = -1$, $a_2 = 2$ up to the second order in $\epsilon$.}
\label{LimCycTab}
\end{center}
\end{table*}
%%%%%%%%%%%%%%%%%%%%%%%%%%%%%%%%%%%%%%%%%
\begin{eqnarray}
f_2^{(1)}(\xi)  &=& - A_0 \omega_1 \Big[ \frac{a_2(1 + \tau_0)}{\omega_0} \sin (\omega_0 \tau_0) \cos \xi \nonumber \\ 
&& \hspace{10mm} + \hspace{1mm} \frac{\tau_0}{a_1\sin (\omega_0 \tau_0)} \sin \xi \Big] \,. 
\end{eqnarray}
Thus, according to the conditions (\ref{Cond1}), (\ref{Cond2}), we must demand
\begin{eqnarray}
- \frac{2 A_0 \omega_1 \tau_0}{a_1 \sin^2(\omega_0 \tau_0)} = 0  \quad \textrm{and} \quad - \frac{4 A_0 \omega_1 (1 + \tau_0)}{a_1 \sin(2 \omega_0 \tau_0)} = 0\, .
\end{eqnarray}
We must thus have either $A_0 = 0$ or $\omega_1 = 0$.  If we choose $A_0 =0$, we only obtain the trivial solution.  Thus, we choose $\omega_1 = 0$, and the coefficient $A_0$ is yet to be determined.  The solution for $\mathbf{V}^{(1)} (\xi)$ is then simply given by the solution to the homogeneous system:
\begin{eqnarray}
V_1^{(1)}(\xi) &=& A_1 \cos \xi , \label{V11} \\ 
V_2^{(1)}(\xi) &=& -\frac{A_1}{a_1 \sin(\omega_0 \tau_0)} \sin \xi \, ,  \label{V21}
\end{eqnarray}
where $A_1$ is to be determined in higher orders.

Expanding (\ref{ddesys1}), (\ref{ddesys2}) up to order $\epsilon^2$ while taking into account the zeroth- and first-order result, we obtain the inhomogeneity $\mathbf{f}^{(2)}(\xi)$ as given by the expansion (\ref{fFouExp}). For the first component we have
{\allowdisplaybreaks
\begin{eqnarray}
\alpha_{1,1}^{(2)} &=& -\frac{A_0^3(1 + \omega_0^2)}{4 a_1^2 \omega_0} + A_0(\omega_0 + \tau_0 \omega_2)\, , \\ 
\beta_{1,1}^{(2)} &=& \frac{A_0^3(1 + \omega_0^2)}{4 a_1^2} + \frac{A_0(1 + \tau_0 ) \omega_2}{\omega_0} + A_0 \, , \\
\alpha_{1,3}^{(2)} &=& \frac{A_0^3(3\omega_0^2 - 1)}{12 a_1^2} \, , \\
\beta_{1,3}^{(2)} &=& \frac{A_0^3(3 - \omega_0^2)}{12 a_1^2} \, .
\end{eqnarray}
And for the second component we have
\begin{eqnarray}
\alpha_{2,1}^{(2)} &=& - a_2 \sin(\omega_0 \tau_0)\left[\frac{A_0^3}{4} + \frac{A_0(1+ \tau_0)\omega_2}{\omega_0} + A_0\right] , \hspace{3mm} \\
\beta_{2,1}^{(2)} &=& -a_2 \sin(\omega_0 \tau_0)\left[\frac{A_0^3}{4 \omega_0} + A_0 (\omega_0 + \tau_0 \omega_2) \right] , \\
\alpha_{2,3}^{(2)} &=&- a_2 \sin(\omega_0 \tau_0)\frac{A_0^3}{12}\left[2 \cos(2 \omega_0 \tau_0) - 1\right] \, , \\
\beta_{2,3}^{(2)} &=&- a_2 \sin(\omega_0 \tau_0)\frac{A_0^3}{12 \omega_0}\left[2 \cos(2 \omega_0 \tau_0) + 1\right] \, ,
\end{eqnarray}
while all other coefficients vanish.
Imposing the conditions (\ref{Cond1}), (\ref{Cond2}) on the inhomogeneity $\mathbf{f}^{(2)}(\xi)$, we obtain the system of equations
}
\begin{eqnarray}
A_0^2(1 + a_1^2 + \omega_0^2) - 8a_1^2\omega_0(\omega_0 + \omega_2 \tau_0) &=& 0 \,, \hspace{3mm} \\
A_0^2\omega_0 (1 + a_1^2 + \omega_0^2) + 8 a_1^2\omega_0 + 8a_1^2 (1 + \tau_0)\omega_2 &=& 0 \, . \hspace{3mm}
\end{eqnarray}
Its solutions read
\begin{eqnarray}
\omega_2 &=& -\frac{\omega_0 + \omega_0^3}{1 + \tau_0 + \tau_0 \omega_0^2} \, ,  \label{om2} \\ 
A_0 &=& \pm \sqrt{\frac{8a_1^2 \omega_0^2}{(1+ a_1^2 + \omega_0^2)(1 + \tau_0 + \omega_0^2 \tau_0)}} \, . \label{A0}
\end{eqnarray}
Choosing the sign of $A_0$ to be positive fixes the phase of our zeroth order solution definitively.  This procedure can easily be carried to higher orders, where only even orders lead to nonzero terms for both the corrections to the angular frequency $\omega_n$ and the expansion of the limit cycle $\mathbf{V}^{(n)}(\xi)$. Expanding (\ref{ddesys1}), (\ref{ddesys2}) to order $\epsilon^{2n}$, we find the coefficient $A_{2(n-1)}$ and the correction $\omega_{2n}$ from the conditions (\ref{Cond1}), (\ref{Cond2}).
%%%%%%%%%%%%%
\begin{figure*}[t]
  \begin{center}
    \epsfig{file=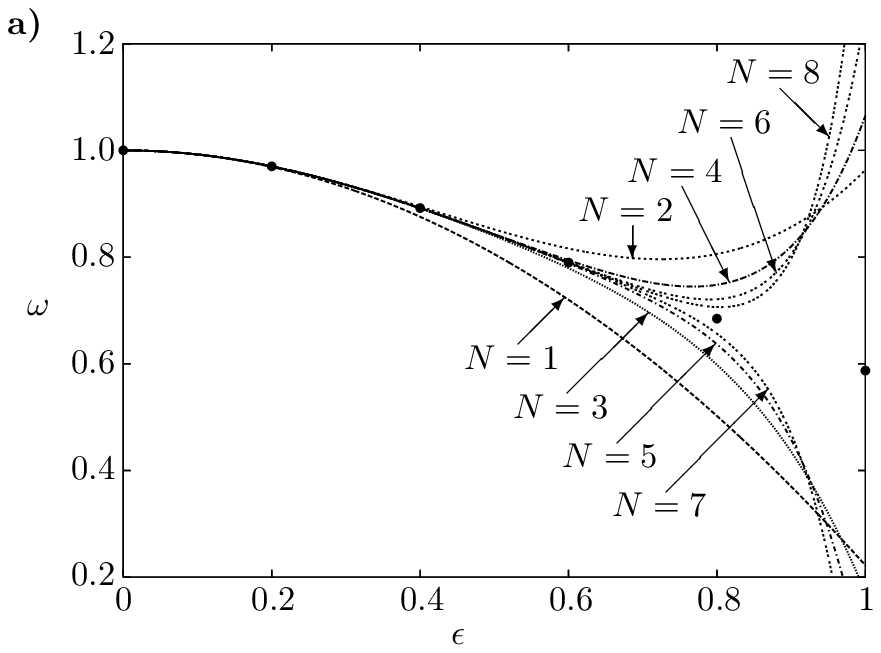,width=\columnwidth} 
    \hfill
    \epsfig{file=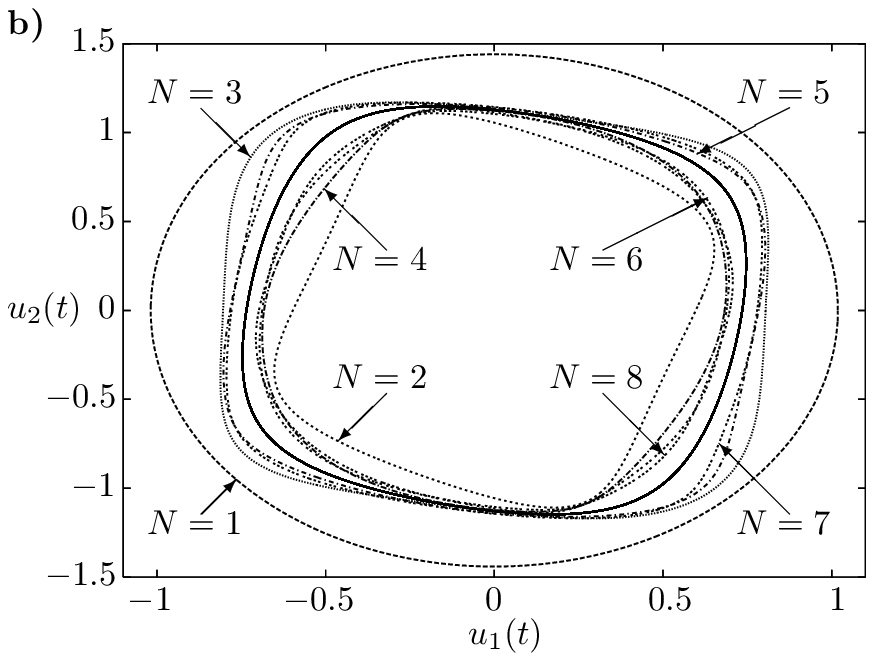,width=\columnwidth}
    \caption{Perturbative results for the angular frequency $\omega$ and the limit cycle $\{u_1(t), \, u_2(t)\}$. In {\bf a)} the angular frequency is shown as a function of $\epsilon$. The dashed lines represent the perturbative results as given by (\ref{OmPer}) and Tab.~\ref{OmTab}. Numerical results are shown by dots. In {\bf b)} the limit cycle $\{u_1(t), \, u_2(t)\}$ is shown for $\epsilon = 1$. Dashed lines represent perturbative results according to (\ref{LimPer}); the numerical result is shown by the solid line.}
\label{fig2}
  \end{center}
\end{figure*}
%%%%%%%%%%%%%%%%%%
 
From here on, we consider the choice of parameters $a_1 = -1$, $a_2 = 2$. 
These parameter values lead to $\omega_0 = 1$, $\tau_0 = \pi/4$ and the solution (\ref{om2}), (\ref{A0}) reduces to
\begin{eqnarray}
\omega_2 = -\frac{4}{2 + \pi}\, ,  \quad \quad A_0 = \frac{4}{\sqrt{3(2 + \pi)}} \, .
\end{eqnarray}
Table \ref{OmTab} shows the first eight nonvanishing corrections to the angular frequency. Note that the signs of the expansion coefficients $\omega_n$ alternate and that their absolute value grows rapidly. This indicates that the perturbation series for $\omega$ is a divergent Borel series. Table \ref{LimCycTab} shows the expansion coefficients of the first two nonvanishing orders of the Fourier expansion of the limit cycle as given by (\ref{VFouExp}). 
Figure \ref{fig2} {\bf a)} shows the first eight orders of the perturbatively calculated angular frequency $\omega^{(N)}$,
\begin{eqnarray}
\omega^{(N)} = \sum_{n=0}^N \omega_{2n} \epsilon^{2n}\,,  \label{OmPer}
\end{eqnarray}
as a function of $\epsilon$. Note that odd and even perturbation orders yield results which are respectively smaller and larger than the numerical values. For small values of the delay, the perturbative solution is in good agreement with the numerical data.  However, as $\epsilon$ grows, the perturbative solution becomes unacceptable. Figure \ref{fig2} {\bf b)} shows an example of the perturbatively calculated limit cycle given by
\begin{eqnarray}
\mathbf{u}^{(N)}(t) = \epsilon \sum_{n=0}^{N-1}\mathbf{V}^{(2n)}(\xi/ \omega)\epsilon^{2n} \, , \label{LimPer}
\end{eqnarray}
where we count the order $N$ of our perturbation expansion such that in the $N$th order we obtain the $N$th nonvanishing corrections $\omega_{2N}$ and $\mathbf{V}^{(2(N-1))}(\xi)$. For the value $\epsilon=1$ chosen in Fig.~\ref{fig2} {\bf b)}, the limit cycle can still be obtained with good precision from the perturbation series (\ref{LimPer}) and as in the case of the angular frequency we observe that the perturbative approximations approach the numerical result in an alternating manner. However, as $\epsilon$ increases, the perturbative solution (\ref{LimPer}) becomes useless as in the case of the angular frequency. 
Thus, if we want to obtain analytical results for larger values of the temporal delay $\tau$, we must resum our perturbation series. In the next section, we apply a Shohat transformation to the perturbative results for both the angular frequency $\omega^{(N)}$ and the limit cycle $\mathbf{u}^{(N)}(t)$.
%%%%%%%%%%%%%%%%%%%%%%%%%%%%%%%%%%%%%%%%%%%%%
\section{Shohat Expansion} \label{SE}
%%%%%%%%%%%%%%%%%%%%%%%%%%%%%%%%%%%%%%%%%%%%%
Now, we resum our perturbative results by performing a Shohat expansion. This method was first introduced for calculating the period of a Van der Pol oscillator in Ref.~\cite{Shohat} and it has been conjectured that the expansion yields results which are valid for all values of the perturbation parameter \cite{Shohat,Bellman}. Furthermore, it has been stated that the Shohat expansion is succesful when the periodic solution to the differential equation in question is of softening type, i.e., the angular frequency $\omega$ decreases with $\epsilon$ \cite{Sarma}, which is the case for our system as is evident from Fig.~\ref{fig2} {\bf a)}.

The basic idea of the Shohat expansion is to map the perturbation parameter $\epsilon \in [0, \hspace{1mm} \infty)$ to a new parameter  $\mu \in [0, \hspace{1mm} 1)$. In order to perform the resummation of the angular frequency, we introduce the new expansion parameter $\mu$ according to the transformation suggested by Shohat in Ref.~\cite{Shohat} and thus set
\begin{eqnarray}
\mu = \frac{\epsilon^2}{1 + \epsilon^2} \, , \label{mu}
\end{eqnarray}
where we explicitly take into account that the perturbation series for the angular frequency depends only on even powers of $\epsilon$. Inverting (\ref{mu}), we have
\begin{eqnarray}
\epsilon^2 = \frac{\mu}{1-\mu} \, . \label{epsmu}
\end{eqnarray}
We now obtain the Shohat expansion of our perturbative result by replacing $\epsilon^2$ in (\ref{OmPer}) according to the last identity and re-expanding the series in $\mu$ up to order $\mu^N$. The Shohat expansion of the angular frequency is thus given by
\begin{eqnarray}
\omega_{\rm S}^{(N)} = \sum_{n = 0}^N \mu^n \sum_{k = 0}^n  \binom{n-1}{k} \omega_{2(n - k)} \, . \label{OmS}
\end{eqnarray}
The resummation of the limit cycle (\ref{LimPer}) is performed in a similar manner and we obtain
\begin{eqnarray}
\mathbf{u}_{\rm S}^{(N)}(t) = \epsilon \sum_{n=0}^{N-1} \mu^n \sum_{k = 0}^n \binom{n-1}{k} \mathbf{V}^{(2(n-k))}(t)\, .  \label{LimS}
\end{eqnarray}
Finally, in order to evaluate the resummed angular frequency and limit cycle for a certain value of $\epsilon$, we replace $\mu$ in (\ref{OmS}) and (\ref{LimS}) according to (\ref{mu}). 
%%%%%%%%%%%%%%%%%%%%%%%%%%%%%%%%%
\begin{figure*}[t]
  \begin{center}
    \epsfig{file=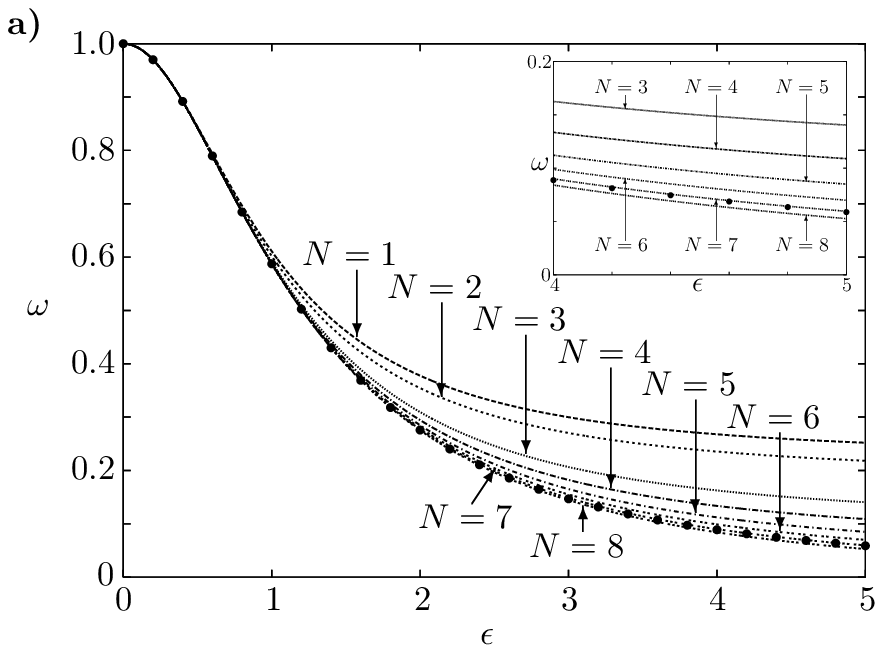,width=\columnwidth} 
    \hfill
    \epsfig{file=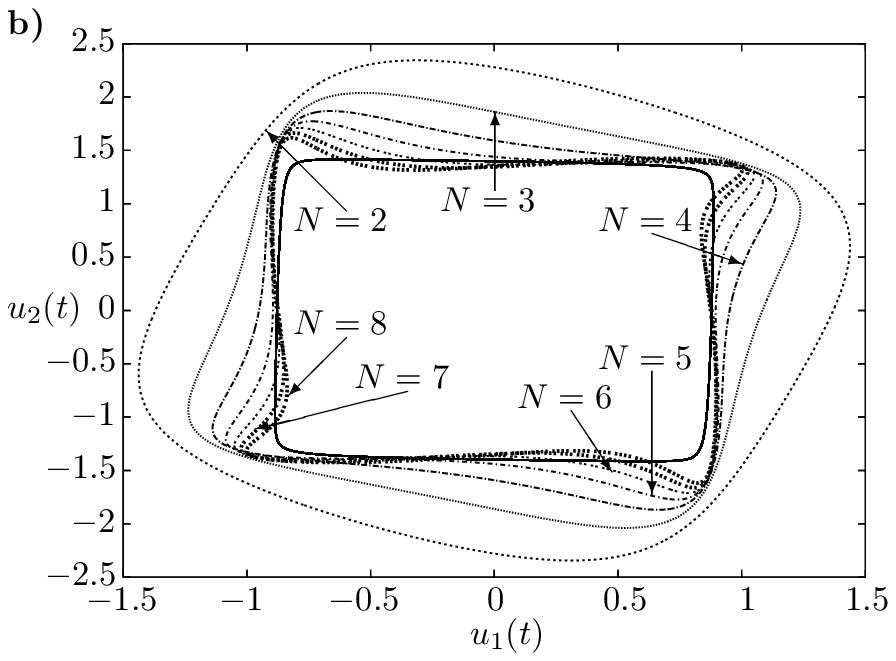,width=\columnwidth}
    \caption{Angular frequency and limit cycle after Shohat resummation. In {\bf a)} the angular frequency is shown as a function of $\epsilon$. Results from the Shohat expansion as given by (\ref{OmS}) are shown by dashed lines. Numerical results are shown by dots. The inset shows a magnification of the interval $4 \leq \epsilon \leq 5$. In {\bf b)} the limit cycle is shown for $\epsilon = 2$. Dashed lines represent results from the Shohat expansion as given by (\ref{LimS}); the numerical result is shown by the solid line.}
\label{fig3}
  \end{center}
\end{figure*}
%%%%%%%%%%%%%%%%%%%%%%%%%%%%

Figure \ref{fig3} {\bf a)} shows the angular frequency after Shohat resummation as a function of the delay parameter $\epsilon$. We find that, if we go to sufficiently high orders, the resummed expansion yields reasonably good results for all values of $\epsilon$. Figure \ref{fig3} {\bf b)} shows an example of the limit cycle after resummation. Note that for the value of the delay parameter in Fig.~\ref{fig3} {\bf b)} the perturbative result prior to resummation would be completely useless. 

\begin{figure*}[t]
  \begin{center}
    \epsfig{file=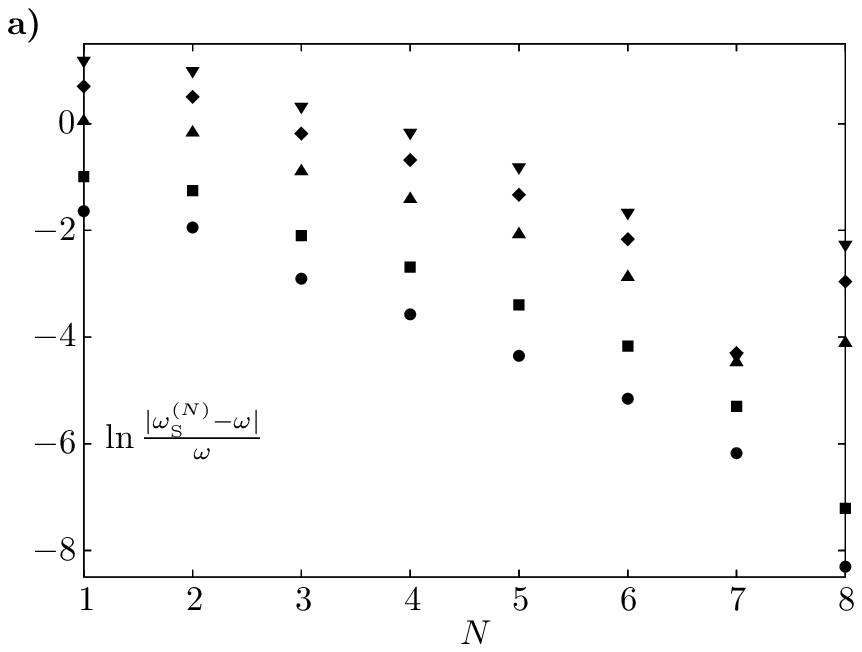,width=\columnwidth} 
        \hfill
    \epsfig{file=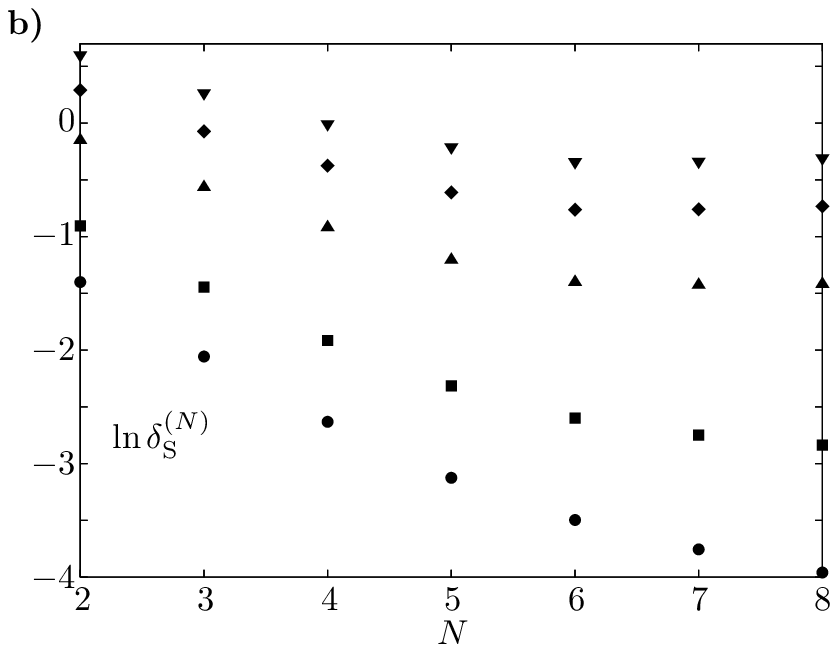,width=\columnwidth} 
    \caption{Convergence of the angular frequency and the limit cycle after Shohat resummation. In {\bf a)} the logarithm of the relative deviation of the angular frequency as given by (\ref{OmS}) from the numerical value and in {\bf b)} the logarithm of the error measure for the limit cycle as given by (\ref{ErMe}) are shown versus the perturbation order. In both {\bf a)} and {\bf b)} different symbols indicate different values of $\epsilon$ (dots: $\epsilon = 1.6$; squares: $\epsilon = 2.0$; triangles: $\epsilon = 3.0$; diamonds: $\epsilon = 4.0$, upside-down triangles: $\epsilon = 5.0$).}
\label{fig4}
  \end{center}
\end{figure*}

Figure \ref{fig4} {\bf a)} shows the convergence of the angular frequency obtained from the Shohat expansion versus the perturbation order $N$ for different values of the temporal delay.  For small values of the delay, the convergence seems to be exponentially fast, at least up to the eighth order. For larger delays, the convergence appears to be less regular. In order to examine the convergence of the limit cycle results, we consider the error measure
\begin{eqnarray}
\delta^{(N)} = \frac{\int_{T_0}^{T_0 + T} dt \hspace{1mm} \left\| \mathbf{u}(t) - \mathbf{u}^{(N)}(t) \right\|_2}{\int_{T_0}^{T_0 + T} dt \hspace{1mm} \left\| \mathbf{u}(t) \right\|_2} \, , \label{ErMe}
\end{eqnarray}
where we rescale the argument of our analytic solution so that its period is identical to the period of the numerical solution and shift the phase of the analytic solution according to the phase of the numerical solution.  Figure \ref{fig4} {\bf b)} shows the convergence of the results for the limit cycle. As in the case of the angular frequency, the results from the Shohat expansion and their convergence with the perturbation order are best as long as the delay is not too large. In the next section, we thus use a more efficient method to resum the perturbation series. It yields accurate results already in low orders, allows us to obtain more precise results, and its convergence depends less crucially on the size of the delay parameter.
%%%%%%%%%%%%%%%%%%%%%%%%%%%%%%%%%%%%%%%%%%%%%%%%%%%%%%%%%
\section{Variational Perturbation Theory} \label{VPT}
%%%%%%%%%%%%%%%%%%%%%%%%%%%%%%%%%%%%%%%%%%%%%%%%%%%%%%%%%
In this section, we improve the resummation of the perturbation series of the angular frequency and the limit cycle by applying VPT to the perturbation series (\ref{OmPer}) and (\ref{LimPer}). This method is based on a variational approach due to Feynman and Kleinert \cite{Feynman2}, which has been systematically extended to the nonperturbative approximation scheme now called VPT \cite{Kleinertsys,PathInt3,VerenaBuch,Festschrift}.
\subsection{Basic Principles}
VPT is capable of converting divergent weak-coupling into convergent
strong-coupling expansions and has been applied successfully in various fields, such as quantum mechanics, quantum statistics, condensed matter physics, and the theory of critical phenomena.  In fact, the most accurate critical exponents
come from this theory \cite{seven},
as verified by recent satellite experiments \cite{LIPA}. First applications of VPT in the field of Markov processes and nonlinear dynamics are found in Refs.~\cite{Putz,Dreger} and Ref.~\cite{Schanz}, respectively.
%%%%%%%%%%%%%%%%
%%%%%%%%%%%%%%%%%%%

The convergence of VPT has been analyzed up to very high orders for the ground-state energy
of the anharmonic oscillator
\begin{eqnarray}
\label{1D}
V(x) = \frac{1}{2}\,\omega^2 x^2 + g x^4
\end{eqnarray}
and was found to be exponentially fast \cite{JankeC1,JankeC2}.
This surprising result has been confirmed later by studying other physical systems and was
proven to hold in general \cite{PathInt3,VerenaBuch}. Furthermore, the
exponential convergence seems to be uniform with respect to other system parameters.
The variational resummation of perturbation series thus yields approximations
which are generically reasonable for all temperatures \cite{Meyer,Weissbach},
space and time coordinates \cite{Michael1,Putz,Dreger},
magnetic field strengths \cite{Michael2}, coupling constants \cite{Bender,Schanz2,EffPot}, spatial dimensions \cite{LargeD}, etc.

VPT permits the evaluation of a divergent series of the form
\begin{eqnarray}
f^{(N)}(g) = \sum_{n = 0}^N a_n g^n \label{PerExp}
\end{eqnarray}
and yields a strong-coupling expansion of the generic form
\begin{eqnarray}
f(g) = g^{p/q}\sum_{m = 0}^M b_m g^{-2m/q}\, . \label{fStrGen}
\end{eqnarray}
Here, $p$ and $q$ are real growth parameters and characterize the strong-coupling behavior.  Introducing a scaling parameter $\kappa$, which is
afterwards set to one, Eq.~(\ref{PerExp}) can be rewritten as
\begin{eqnarray}
f^{(N)}(g) = \left. \kappa^p \sum_{n = 0}^N a_n \left(\frac{g}{\kappa^q}\right)^n \right|_{\kappa = 1}\, . \label{PerExpRe}
\end{eqnarray}
Applying Kleinert's square-root trick \cite{PathInt3},
i.e.\ setting 
\begin{eqnarray}
\kappa = K \sqrt{1 + gr} \label{kappaK}
\end{eqnarray}
with
\begin{eqnarray}
r = \frac{\kappa^2 - K^2}{gK^2} \label{rkappa} 
\end{eqnarray}
in (\ref{PerExpRe}), the variational parameter $K$ is introduced into the
perturbation series:
\begin{eqnarray}
f^{(N)}(g) = \sum_{n = 0}^N a_n g^n K^{p - nq} \hspace{1mm} (1 + gr)^{(p -nq)/2} \Big|_{\kappa
  = 1} \, .
\end{eqnarray}
The Taylor series of $(1 + gr)^{\alpha}$ with $\alpha \equiv (p - nq)/2$ reads
\begin{eqnarray}
(1 + gr)^{\alpha} \Big|_{\kappa = 1} = \sum_{k = 0}^{N - n} {\alpha \choose k} \left( \frac{1}{K^2} -1 \right)^k + {\cal O} \left(g^{N-n+1}\right) \,, \hspace{-4mm} \nonumber \\ \label{Tayfact}
\end{eqnarray}
where the generalized binomial coefficient is defined by
\begin{eqnarray}
{\alpha \choose k} \equiv \frac{\Gamma (\alpha + 1)}{\Gamma (k + 1)\Gamma (\alpha - k + 1)}  \, .
\end{eqnarray}
The series (\ref{Tayfact}) is truncated after $k = N-n$ since the original
function $f^{(N)}(g)$ is only known up to order $g^N$.  As a result of
this truncation, the function $f^{(N)}(g)$ becomes dependent on the
variational parameter $K$:
\begin{eqnarray}
f^{(N)}(g, K) &=&  \label{fNgK} \\
&& \hspace{-12mm} \sum_{n = 0}^{N} a_n g^n K^{p - nq} \sum_{k = 0}^{N - n} \binom{(p - nq)/2}{k} 
\left( \frac{1}{K^2} - 1 \right)^k \,.  \nonumber
\end{eqnarray}
The influence of the variational parameter is then optimized according to the {\it principle of minimal sensitivity} \cite{Stevenson}, i.e., one evaluates the function (\ref{fNgK}) at that value of the variational parameter $K$ for which it has an extremum or turning point. In the following, we set $g = \epsilon^2$ in (\ref{OmPer}) and (\ref{LimPer}).
%%%%%%%%%%%%%%%%%%%%%%%%%%%%%%%%%%%%%%%%%%%%%%%%%%%%%%%
\subsection{Resummation of the Angular Frequency}
\begin{figure*}[t]
  \begin{center}
    \epsfig{file=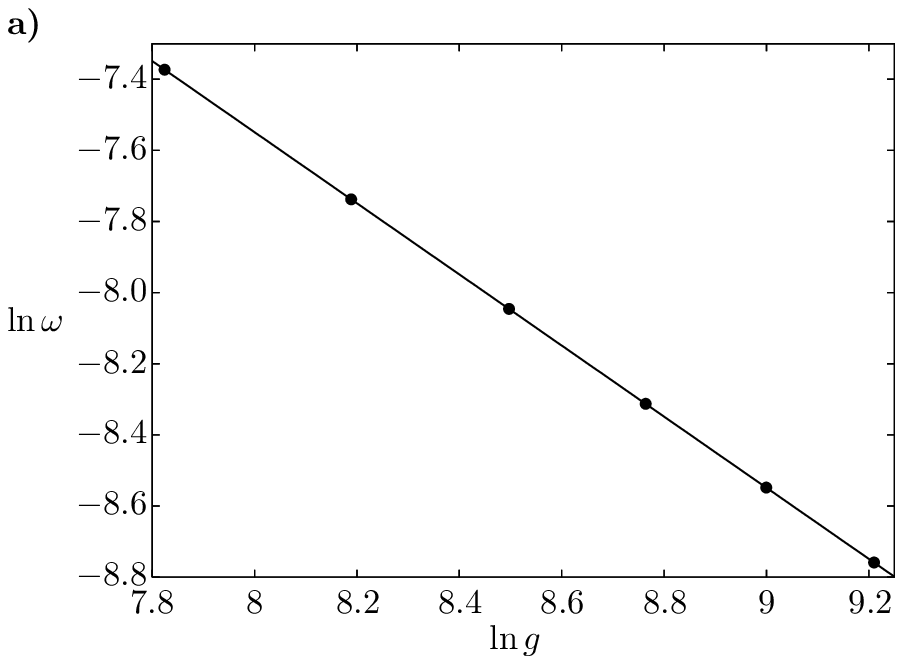,width=\columnwidth} 
        \hfill
    \epsfig{file=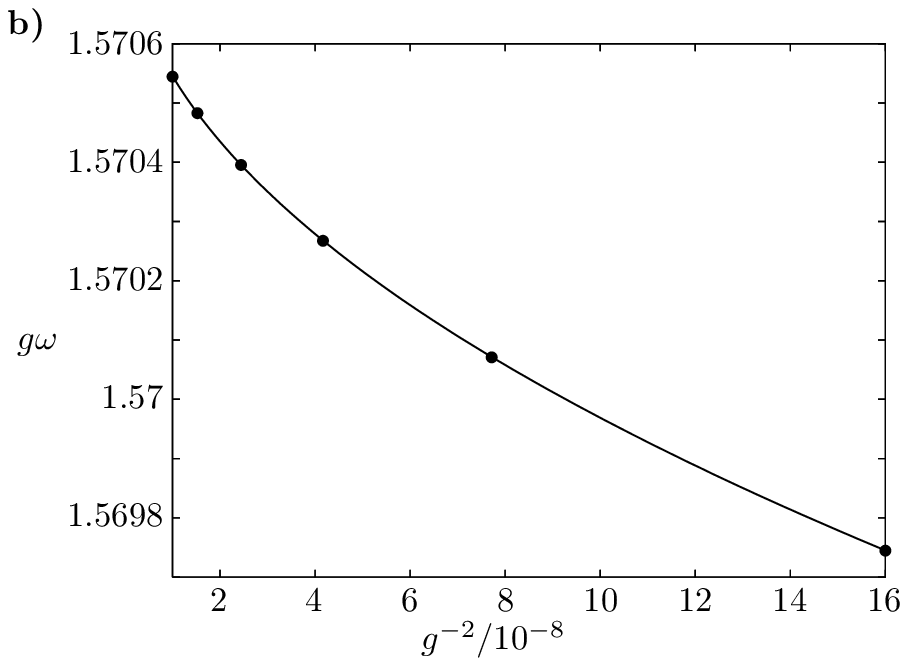,width=\columnwidth} 
    \caption{Angular frequency for large delays ($\sqrt{\tau - \tau_0} \in [50, \, 100]$). In {\bf a)} the logarithm of the angular frequency is shown versus the logarithm of the delay parameter $g$.  Numerical data are represented by dots; the solid line represents a fit of the data to a function of the form (\ref{fit1}). In {\bf b)} the product of the delay parameter and the angular frequency is shown versus the inverse square of the delay parameter. Numerical data are represented by dots; the solid line represents a fit of the data to a function of the form (\ref{fit2}).}
\label{fig5}
  \end{center}
\end{figure*}

%%%%%%%%%%%%%%%%%%%%%%%%%%%%%%%%%%%%%%%%%%%%%%%%%%%%%%%
We now apply VPT to obtain an improved resummation of the angular frequency (\ref{OmPer}). VPT is applicable when the physical quantity in question has a strong-coupling expansion of the form (\ref{fStrGen}) \cite{PathInt3,VerenaBuch}. Therefore, we first consider our numerical data for the angular frequency in the case of large delays and determine the growth parameter $p$ and $q$ in (\ref{fStrGen}). To this end, we analyze our numerical data in two steps. First, in Fig.~\ref{fig5}~{\bf a)}, we plot our numerical results for $\ln \omega$ versus $\ln g = \ln (\tau - \tau_0)$.  Fitting our data  to a function of the form 
\begin{eqnarray}
f(\ln g) = p/q \ln g + \ln b_0 \, , \label{fit1}
\end{eqnarray} 
we find $p/q = -0.9997$ and $b_0 = 1.565$. We expect the growth parameters to be integers and thus set $p/q = -1$. For large delays, the leading asymptotic behavior of $\omega$ is thus given by $\omega \sim g^{-1}$.  In order to determine not only the ratio of $p$ to $q$ but the individual values of the growth parameters, we then fit our data for $g \omega$ to a function of the form 
\begin{eqnarray}
f(g^{-2}) = b_0 + b_1 g^{-2/q}\, ,\label{fit2}
\end{eqnarray} 
which is shown in Fig.~\ref{fig5}~{\bf b)}. The numerical results from the fit are: $b_0 = 1.571$, $b_1 = -2.7$, and $q = 1.993$.  Thus, we assume $q = 2$ and from our previous result we then have $p = -2$.  In order to determine $b_0$ and $b_1$ numerically with better accuracy, we can now perform a hierarchy of approximations to order $M$ by fitting $g \omega $ to functions of the form 
%%%%%%%%%%%%%%%%
%%%%%%%%%%%%%%%%%%%
\begin{eqnarray}
f(g^{-1}) = \sum_{m = 0}^M b_m g^{-m}\, .
\end{eqnarray} 
From this procedure we obtain the more precise numerical values $b_0 = 1.57081$ and $b_1 = -2.66$. Now, we can introduce the variational parameter $K$ to the perturbation series (\ref{OmPer}) according to (\ref{fNgK}) with $p = -2$, $q = 2$:
\begin{eqnarray}
\omega_{\rm VPT}^{(N)}(g, K) &=& \label{OmVPT}  \\
&& \hspace{-14mm} \sum_{n=0}^N \omega_{2n}g^n K^{p - nq} \sum_{k = 0}^{N - n} \binom{(p - nq)/2}{k} \left( \frac{1}{K^2} - 1 \right)^k \,.  \nonumber
\end{eqnarray}
To first order we obtain
\begin{eqnarray}
\omega_{\rm VPT}^{(1)}(g,K) = \frac{(2 + \pi)(2K^2 - 1) - 4g}{K^4(2 + \pi)} \, , \label{OmVPT1}
\end{eqnarray}
which has a minimum at
\begin{eqnarray}
K^{(1)} = \sqrt{1 + \frac{4g}{2 + \pi}} \, . \label{K1}
\end{eqnarray}
Evaluating (\ref{OmVPT1}) at the optimized value of the variational parameter then yields 
\begin{eqnarray}
\omega^{(1)}_{\rm VPT}(g,K^{(1)}) = \frac{2 + \pi}{4g + 2 + \pi}\, .
\end{eqnarray}
In the limit of large delays, $g \to \infty$, we thus have
%%%%%%%%%%%%%%%%
\begin{figure*}[t]
  \begin{center}
    \epsfig{file=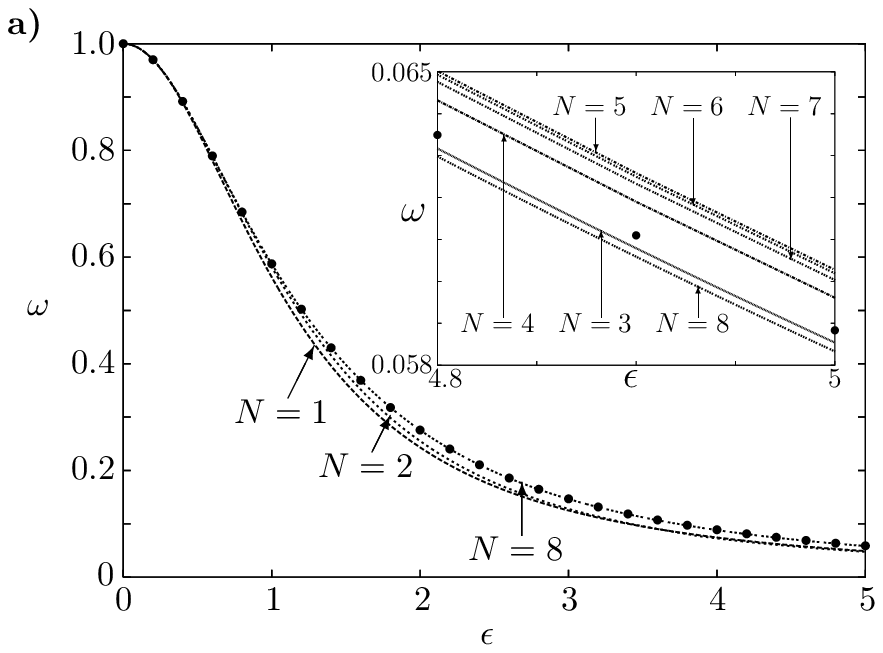,width=\columnwidth} 
        \hfill
    \epsfig{file=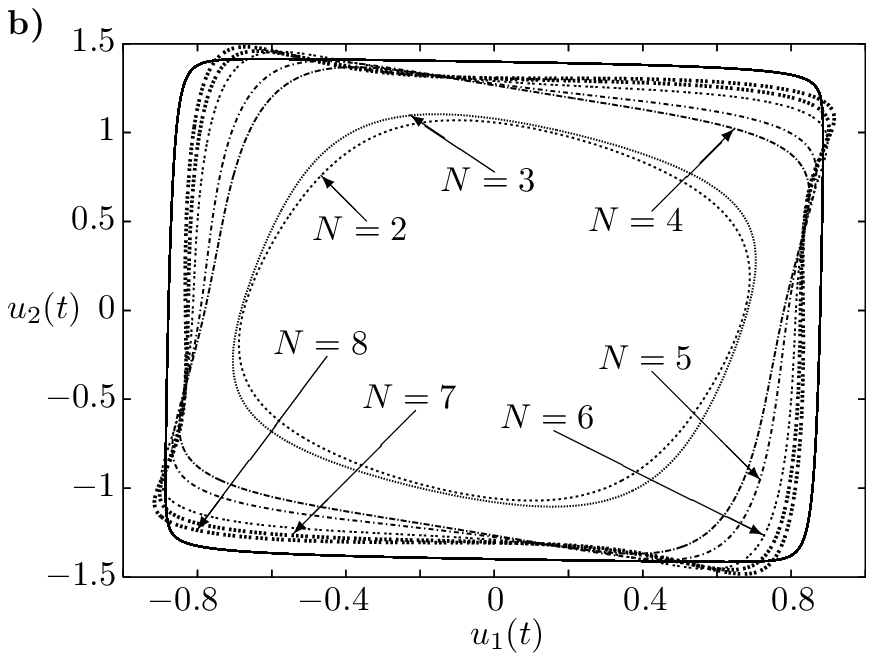,width=\columnwidth} 
    \caption{Angular frequency and limit cycle from VPT. In {\bf a)} the angular frequency  as given by (\ref{OmOpt}) is shown as a function of $\epsilon$ for the orders $N=1$, $2$, $8$ of VPT (orders three through seven would lie very close to the curve for $N=8$). Dots represent numerical values. The inset shows a magnification of the interval $4.8 \leq \epsilon \leq 5$. In {\bf b)} the limit cycle  is shown for $\epsilon = 2$.  Dashed lines represent the results from VPT as given by (\ref{VVPT}). The numerical result is shown by the solid line.}
\label{fig6}    
\end{center}
\end{figure*}
%%%%%%%%%%%%%%%%%%%
\begin{eqnarray}
\hspace{-1mm} \omega^{(1)}_{\rm VPT}(g,K^{(1)})  \sim b_0^{(1)} g^{-1} + b_1^{(1)}g^{-2} \, ,
\end{eqnarray}
with
\begin{eqnarray}
b_0^{(1)} = \frac{2 + \pi}{4} \approx 1.28540
\end{eqnarray}
and
\begin{eqnarray}
b_1^{(1)} = -\frac{(2+\pi)^2}{16} \approx -1.6522 \, .
\end{eqnarray}
To second order, Eq.~(\ref{OmVPT}) yields
%{\allowdisplaybreaks
\begin{eqnarray}
\omega^{(2)}_{\rm VPT}(g,K) &=& \frac{1}{27K^6}\bigg[  27 \left(3 K^4-3 K^2+1 \right)  \label{OmVPT2} \\ && + \hspace{1mm} g\frac{108 \left(2-3 K^2\right) }{2+\pi } + g^2\frac{4 (341+108 \pi ) }{(2+\pi )^3} \bigg]\, . \nonumber 
\end{eqnarray}
%}
Since this has no real extremum in the variational parameter $K$, we look for roots of the second derivative.  

In general, in order to optimize the influence of the variational parameter, we first look for minima or maxima of $\omega_{\rm VPT}^{(N)}(g,K)$, and if those do not exist, for positive roots of higher derivatives.  In each order $N$, the optimized variational parameter $K^{(N)}$ is thus determined from the condition
\begin{eqnarray}
\left. \frac{d \omega_{\rm VPT}^{(N)}(g,K)}{d K} \right|_{K = K^{(N)}} &=& 0 \nonumber \\ {\rm or} \quad \left. \frac{d^2 \omega_{\rm VPT}^{(N)}(g,K)}{d^2 K} \right|_{K = K^{(N)}} &=& 0 \,, \, \ldots  \, . \label{KCond}
\end{eqnarray}
In cases where a certain derivative has several positive roots, we choose the one which is closest to the optimized value from the previous order $K^{(N-1)}$.  The $N$th order VPT approximation of the angular frequency is then obtained by evaluating (\ref{OmVPT}) for the value of the optimized variational parameter:
\begin{eqnarray}
\omega^{(N)}_{\rm VPT}(g) = \omega_{\rm VPT}^{(N)}(g, K^{(N)}) \, . \label{OmOpt}
\end{eqnarray}

Returning to (\ref{OmVPT2}), we find that for
\begin{eqnarray}
g \leq \frac{3 (2+\pi ) [24+12 \pi +5 \sqrt{35 (2+\pi )}]}{587 - 144 \pi} \approx 14.756 \label{gCond}
\end{eqnarray}
$\omega_{\rm VPT}^{(2)}(g,K)$ has two positive turning points:
\begin{eqnarray}
\tilde{K}^{(2)}_{\pm}  = \frac{\sqrt{60 +15 \pi (4+ \pi ) + 60 (2+\pi ) g  \pm  2 \eta }}{3 (2+\pi )}  \,  , \label{K2}
\end{eqnarray}
with the abbreviation
\begin{eqnarray}
\eta &=& \sqrt{2+\pi} \\
&& \times \hspace{1mm} \sqrt{9 (2+\pi )^3 + 72 (2+\pi )^2 g - (587 - 144 \pi ) g^2 } \, . \nonumber
\end{eqnarray}
Comparing (\ref{K2}) to (\ref{K1}), we find that $\tilde{K}^{(2)}_{-}$ is closer to $K^{(1)}$ and thus evaluate (\ref{OmVPT2}) for $K=\tilde{K}^{(2)}_{-}$ to obtain
\begin{eqnarray}
\omega_{\rm VPT}^{(2)}(g,\tilde{K}^{(2)}_{-}) &=& \frac{27 (2+\pi )^2}{[15 (2+\pi ) (2 + \pi + 4 g) - 2 \eta]^3} \\ && \hspace{-25mm} \times \hspace{1mm} \Big\{  4 \eta^2 - 42 (2+\pi )(2 + \pi + 4 g)\eta  + (2+\pi )  \nonumber \\ && \hspace{-24mm} \times \hspace{1mm} [117 (2+\pi)^3 + 936 (2+\pi )^2 g + 4 (1061+468 \pi ) g^2 ]\Big \}\, . \nonumber
\end{eqnarray}
However, for delay parameters exceeding the value of $g$ given in (\ref{gCond}), we cannot use $\tilde{K}^{(2)}_{\pm}$ since in this case $\eta$ becomes imaginary. Thus, if we want to consider the limit of large delays, we must optimize the variational parameter by considering the third derivative of $\omega_{\rm VPT}^{(2)}(g,K)$, which turns out to have two positive roots for all positive $g$:
%%%%%%%%%%%%%%%%
\begin{figure*}[t]
  \begin{center}
    \epsfig{file=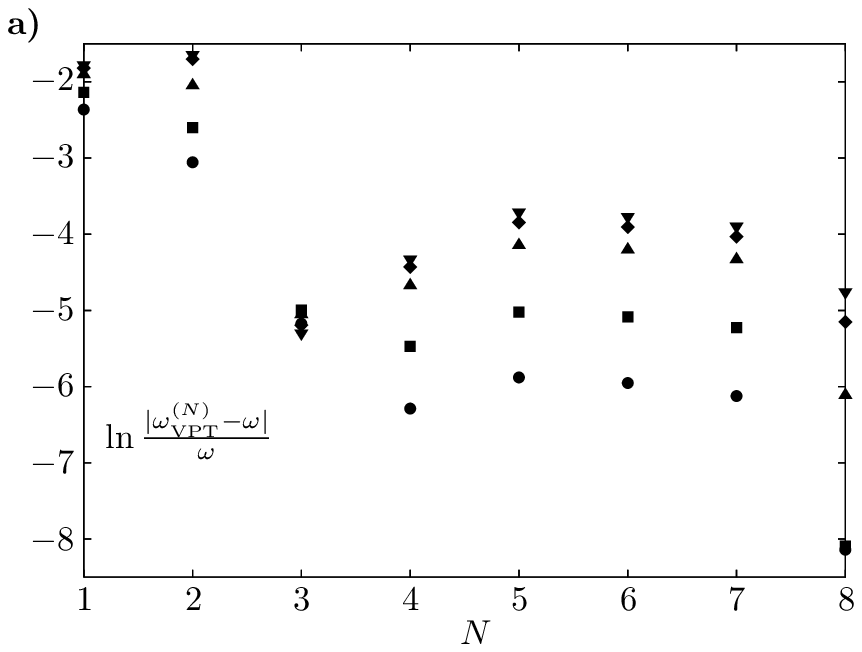,width=\columnwidth} 
        \hfill
    \epsfig{file=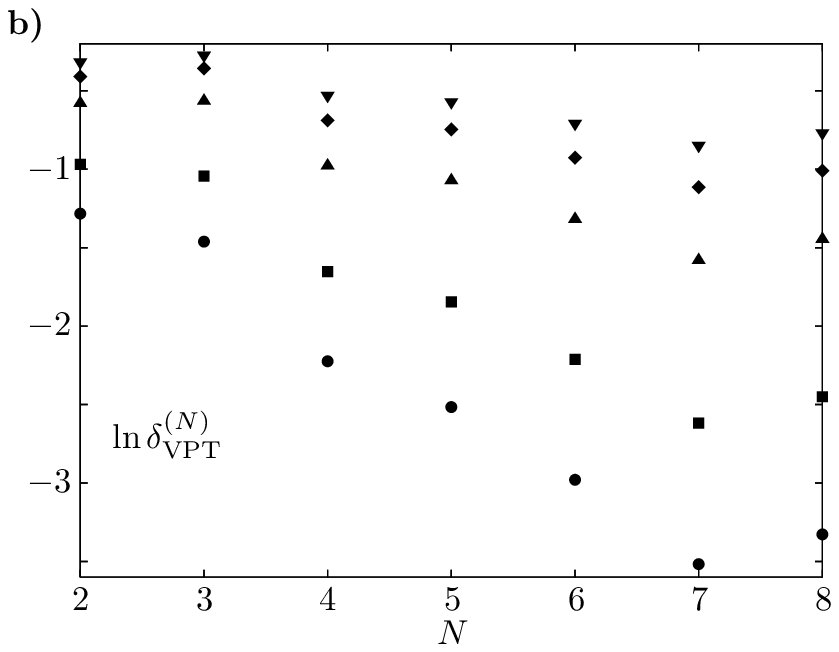,width=\columnwidth} 
    \caption{Convergence of the angular frequency and the limit cycle after resummation with VPT. In {\bf a)} the logarithm of the relative deviation of the angular frequency as given by (\ref{OmOpt}) from the numerical values and in {\bf b)} the logarithm of the error measure for the limit cycle as given by (\ref{ErMe}) are shown versus the perturbation order. In both {\bf a)} and {\bf b)} different symbols indicate different values of $\epsilon$ (dots: $\epsilon = 1.6$; squares: $\epsilon = 2.0$; triangles: $\epsilon = 3.0$; diamonds: $\epsilon = 4.0$, upside-down triangles: $\epsilon = 5.0$).}
\label{fig7}
  \end{center}
\end{figure*}
%%%%%%%%%%%%%%%%%%%
\begin{eqnarray}
K^{(2)}_{\pm} = \frac{\sqrt{180 + 45 \pi( 4 + \pi) + 180(2 + \pi) g \pm \rho}}{3 \sqrt{2}(2 + \pi)}\, , 
\end{eqnarray}
with the abbreviation
\begin{eqnarray}
\rho &=& \sqrt{2+ \pi} \\
&& \hspace{-8mm} \times \hspace{1mm} \sqrt{513(2 + \pi)^3 + 4104(2+ \pi)^2 g + 16(513 \pi - 724)g^2}\, . \nonumber 
\end{eqnarray}
Again, $K^{(2)}_{-}$ is closer to the first-order solution, and we set $K^{(2)} = K^{(2)}_{-}$, to obtain
\begin{eqnarray}
\omega_{\rm VPT}^{(2)}(g,K^{(2)}) &=& \frac{54 (2+\pi )^2}{[45 (2+\pi ) (2 + \pi + 4 g) - \rho]^3}  \nonumber \\ && \hspace{-28mm} \times \hspace{1mm} \Big\{  \rho^2 - 72 (2+\pi )(2 + \pi + 4 g)\rho  + (2+\pi )[1323 (2+\pi)^3 \nonumber \\ && \hspace{-28mm} + \hspace{1mm} 10584 (2+\pi )^2 g + 16(2771+1323 \pi ) g^2 ]\Big \}\, .
\end{eqnarray}
Expanding the last result in $g^{-1}$, we obtain
\begin{eqnarray}
\omega^{(2)}_{\rm VPT}(g,K^{(2)}) &\sim& b_0^{(2)}g^{-1} + b_1^{(2)}g^{-2} \, ,
\end{eqnarray}
with
\begin{eqnarray}b_0^{(2)} &=&
\frac{27 (2+\pi )^3}{2 \left[90+45 \pi - \sqrt{(2+\pi ) (513 \pi - 724)}\right]^3} 
\\ &&  \hspace{-12mm} \times \hspace{0.2mm} \left[2047+1836 \pi -72 \sqrt{(2+\pi ) (513 \pi -724)}\right] \approx 1.23174 \nonumber \, 
\end{eqnarray}
and
\begin{eqnarray}
b_1^{(2)} &=&
\frac{243 (2+\pi )^5}{8 \left[90+45 \pi - \sqrt{(2+\pi ) (513 \pi - 724)}\right]^4} 
\\ &&  \hspace{-12mm} \times \hspace{-0.5mm} \Bigg\{ \hspace{-0.5mm} 63213 \sqrt{\frac{2+ \pi}{513 \pi - 724}} + 162 \pi \left[437\sqrt{\frac{2 + \pi}{513 \pi - 724}} - 82\right] \nonumber \\
&& \hspace{-12mm} +  \hspace{1mm} 426 \sqrt{(2 + \pi)(513 \pi - 724)} - 16193  \Bigg\} \approx -1.1229 \nonumber \, .
\end{eqnarray}
It thus turns out that the second order approximation for the leading and subleading large-delay coefficient is actually worse than the first order one.  However, the results in higher orders turn out to be improved approximations.  For fixed values of the coupling constant, the procedure in higher orders is analogous to the first and second order, where the roots of the first, second, or third derivative of $\omega_{\rm VPT}^{(N)}(g, K)$ have to be determined numerically. Furthermore, in order to obtain the coefficients $b_0^{(N)}$ and $b_1^{(N)}$, we expand the derivatives of $\omega_{\rm VPT}^{(N)}(g, K)$ in $g^{-1}$ and the variational parameter $K$ as
\begin{eqnarray}
K^{(N)} = K_0^{(N)} g^{1/2} + K_1^{(N)}g^{-1/2} + \ldots 
\end{eqnarray}
in order to carry out the optimization procedure.

\begin{figure*}[t]
  \begin{center}
    \epsfig{file=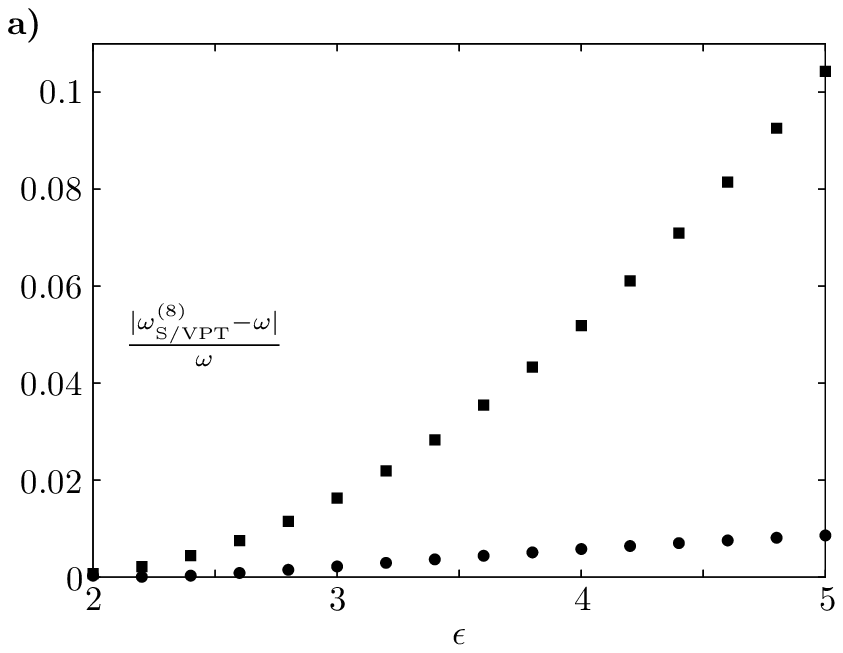,width=\columnwidth} 
        \hfill
    \epsfig{file=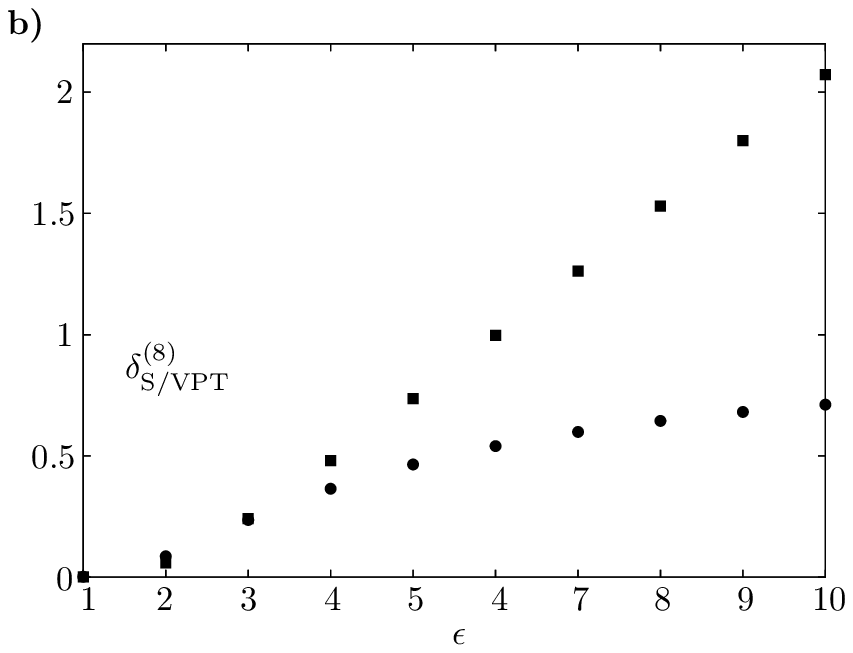,width=\columnwidth} 
    \caption{Comparison of the eighth order results for {\bf a)} the angular frequency and {\bf b)} the limit cycle obtained from the Shohat expansion and VPT. The relative deviations of the analytical results from the corresponding numerical values are shown versus the delay parameter (Shohat expansion: squares; VPT: circles).}
\label{fig8}
  \end{center}
\end{figure*}

Fig.~\ref{fig6} {\bf a)} shows our VPT results for the angular frequency versus the delay parameter $\epsilon$.  The first order result is already in good agreement with the numerical results for a wide range of delays and is far superior to the first order result from the Shohat expansion (compare Fig.~\ref{fig3} {\bf a)}).   Figure \ref{fig7} {\bf a)} shows the convergence of our VPT results for five different values of the delay. The accuracy of our VPT results improves with increasing order; however, not as regularly as in the case of the Shohat expansion for small delays. Figure~\ref{fig8} {\bf a)} shows a comparison of the eighth order results obtained from Shohat resummation and VPT.  In particular, for larger values of the delay, the results from VPT are far superior to the ones from Shohat resummation. Table \ref{LarDelTab} shows our results for the leading large-delay coefficients $b_0$ and the subleading coefficient $b_1$; again, the convergence is not montonic, but we do observe a general trend towards improved results in higher orders. 
%%%%%%%%%%%%%%%%
%%%%%%%%%%%%%%%%%%%
%%%%%%%%%%%%%%%%%%%%%%%%%%%%%%%%%%%%%%%%%%%%%%%%%%%%%%%
\subsection{Resummation of the Limit Cycle}
%%%%%%%%%%%%%%%%%%%%%%%%%%%%%%%%%%%%%%%%%%%%%%%%%%%%%%%
We now proceed to perform a variational resummation of the limit cycle following the approach of Ref.~\cite{Kleinetkathoefer}.  To this end, we consider the perturbation series of each coefficient in the Fourier expansion of $\mathbf{V}(\xi)$ as given by (\ref{VFouExp})
\begin{eqnarray}
A_{1/2, k}^{(N)} &=& \sum_{n = 0}^{N-1}a_{1/2,k}^{(2n)}g^n\,, \label{A12Nk} \\
B_{1/2, k}^{(N)} &=& \sum_{n = 0}^{N-1}b_{1/2,k}^{(2n)}g^n\,. \label{B12Nk}
\end{eqnarray} 
We introduce the variational parameter $K$ 
into the perturbation series for $A_{1/2, k}^{(N)}$ and $B_{1/2, k}^{(N)}$in the same way as for the angular frequency, and obtain by applying (\ref{fNgK}) to the Fourier expansions (\ref{A12Nk}), (\ref{B12Nk})
\begin{eqnarray}
A_{1/2,k,{\rm VPT}}^{(N)}(g, K) &=& \label{aVPT}  \\
&& \hspace{-30mm} \sum_{n=0}^{N-1} a_{1/2,k}^{(2n)}g^n K^{p - nq} \sum_{k = 0}^{N - n} \binom{(p - nq)/2}{k} \left( \frac{1}{K^2} - 1 \right)^k  \nonumber
\end{eqnarray}
and
\begin{eqnarray}
B_{1/2,k,{\rm VPT}}^{(N)}(g, K) &=& \label{bVPT}  \\
&& \hspace{-30mm} \sum_{n=0}^{N-1} b_{1/2,k}^{(2n)}g^n K^{p - nq} \sum_{k = 0}^{N - n} \binom{(p - nq)/2}{k} \left( \frac{1}{K^2} - 1 \right)^k \,.  \nonumber
\end{eqnarray}
Instead of optimizing (\ref{aVPT}) and (\ref{bVPT}) according to the principle of minimal sensitivity, we obtain our VPT result for the limit cycle more easily by evaluating all Fourier expansion coefficients for that value of the variational parameter $K$ which was determined through the optimization procedure of the frequency, i.e., our VPT result for the limit cycle reads:
\begin{eqnarray}
\left(
  \begin{array}{ c }
     V_{1,{\rm VPT}}^{(N)}(\xi) \\
     V_{2,{\rm VPT}}^{(N)}(\xi)
  \end{array}
\right)
& = & \hspace{-1mm}\sum_{k = 1}^{\infty}\Bigg[
\left(
   \begin{array}{ c }
     A_{1, k,{\rm VPT}}^{(N)}(g, K^{(N-1)}) \\
     A_{2, k,{\rm VPT}}^{(N)}(g, K^{(N-1)})
  \end{array}
\right)
\cos k \xi \nonumber
\\ &&
\hspace{-8mm} + \hspace{1mm}
\left(  
\begin{array}{ c }
     B_{1, k,{\rm VPT}}^{(N)}(g, K^{(N-1)}) \\
     B_{2, k,{\rm VPT}}^{(N)}(g, K^{(N-1)})
\end{array}
\right) 
\sin k \xi
\Bigg]\, , \label{VVPT}
\end{eqnarray}
where $K^{(N-1)}$ is determined from the condition (\ref{KCond}) and we use $K^{(N-1)}$ instead of $K^{(N)}$, since the $N$th term in the series for $\mathbf{V}(\xi)$ is a correction of order $g^{N-1}$.

As an example, we consider the lowest order in which we can perform the VPT resummation of the limit cycle. To order $g$ our solution for $\mathbf{V(\xi)}$ reads
%%%%%%%%%%%%%%%%%%%%%%%%%%%%%%%%%%%%%%%%%%%%%%%%%%%%%%%%%%%%%%
\begin{table*}[t]
\begin{center}
\begin{tabular}{|c|c|c|c|c|c|c|c|c||c|}
\hline
\rule[-1mm]{0mm}{7mm}\hspace*{1mm} $N$ \hspace*{1mm} & \hspace*{1mm} $1$ \hspace*{1mm} & \hspace*{1mm} $2$ \hspace*{1mm} & \hspace*{1mm} $3$ \hspace*{1mm} & \hspace*{1mm} $4$ \hspace*{1mm} & \hspace*{1mm} $5$ \hspace*{1mm} & \hspace*{1mm} $6$ \hspace*{1mm} & \hspace*{1mm} $7$ \hspace*{1mm} & \hspace*{1mm} $8$ \hspace*{1mm} & \hspace*{1mm} numerical  \hspace*{1mm} \\[1mm] \hline
\rule[-1mm]{0mm}{7mm}\hspace*{1mm} $b_0^{(N)}$ \hspace*{1mm} & \hspace*{1mm} $1.2854 $ \hspace*{1mm} & \hspace*{1mm} $ 1.23174$ \hspace*{1mm} & \hspace*{1mm} $ 1.56495$ \hspace*{1mm} & \hspace*{1mm} $ 1.59507$ \hspace*{1mm} & \hspace*{1mm} $ 1.61990$ \hspace*{1mm} & \hspace*{1mm} $1.61806$ \hspace*{1mm} & \hspace*{1mm} $ 1.61139$ \hspace*{1mm} & \hspace*{1mm} $ 1.54478$ \hspace*{1mm} & \hspace*{1mm} $ 1.57081 $ \\[1mm] \hline
\rule[-1mm]{0mm}{7mm}\hspace*{1mm} $b_1^{(N)}$ \hspace*{1mm} & \hspace*{1mm} $-1.65$ \hspace*{1mm} & \hspace*{1mm} $-1.12 $ \hspace*{1mm} & \hspace*{1mm} $-2.72 $ \hspace*{1mm} & \hspace*{1mm} $-2.79 $ \hspace*{1mm} & \hspace*{1mm} $-3.05 $ \hspace*{1mm} & \hspace*{1mm} $-3.03 $ \hspace*{1mm} & \hspace*{1mm} $-2.98 $ \hspace*{1mm} & \hspace*{1mm} $-2.21 $ \hspace*{1mm} & \hspace*{1mm} $ -2.66$ \\[1mm] \hline
\end{tabular}
\caption{Leading and subleading coefficients for the large-delay behavior of the angular frequency.}
\label{LarDelTab}
\end{center}
\end{table*}
%%%%%%%%%%%%%%%%%%%%%%%%%%%%%%%%%%%%%%%%%%%%%%%%%%%%%%%%%%
\begin{eqnarray}
V_1(\xi) &=& \frac{4 \cos \xi }{\sqrt{3 (2+\pi )}} - g \bigg\{\frac{5 \sqrt{3}(116+33 \pi ) \cos \xi }{81 (2+\pi )^{5/2}} \\ && + \hspace{1mm}\frac{2 \sqrt{3} }{27 (2+\pi )^{3/2}}[\cos 3 \xi  - 7 \sin 3 \xi ] \bigg \} + {\cal O}(g^2) \, , \nonumber 
\\
%\end{eqnarray}
%\begin{eqnarray}
V_2(\xi) &=& \frac{4 \sqrt{2} \sin \xi }{ \sqrt{3 (2+\pi )}} - g \bigg\{
\frac{ \sqrt{6} (436+93 \pi ) \sin \xi }{81(2+\pi )^{5/2}} \\ &&
- \hspace{1mm} \frac{2 \sqrt{6} }{27 (2+\pi )^{3/2}}[5 \cos 3 \xi  -  \sin 3 \xi] \bigg \}  + {\cal O}(g^2) \, . \nonumber 
\end{eqnarray}
Introducing the variational parameter $K$ according to (\ref{aVPT}), (\ref{bVPT}), we obtain
\begin{eqnarray}
V_{1, {\rm VPT}}^{(2)}(\xi, K) &=& \frac{4(2K^2 - 1) \cos \xi }{K^4 \sqrt{3 (2+\pi )}}  - \frac{g}{K^4} \label{V1VPTOrd1} \\ && \hspace{-26.5mm} \times \hspace{-0.5mm}\bigg\{  \hspace{-0.8mm} \frac{5 \sqrt{3}(116 \hspace{-0.5mm} + \hspace{-0.5mm} 33 \pi ) }{81 (2 \hspace{-0.5mm}+ \hspace{-0.5mm} \pi )^{5/2}} \cos \xi \hspace{-0.4mm}  + \hspace{-0.4mm} \frac{2 \sqrt{3} }{27 (2\hspace{-0.5mm} + \hspace{-0.5mm} \pi )^{3/2}}[\cos 3 \xi \hspace{-0.5mm} -\hspace{-0.5mm} 7 \sin 3 \xi ] \hspace{-0.5mm} \bigg \} ,  \nonumber 
\end{eqnarray}
%\\
\begin{eqnarray}
V_{2, {\rm VPT}}^{(2)}(\xi, K) &=& \frac{4 (2K^2 - 1)\sqrt{2} \sin \xi }{ K^4 \sqrt{3 (2+\pi )}} - \frac{g}{K^4}  \label{V2VPTOrd2}
\\  && \hspace{-26.5mm} \times \hspace{-0.5mm} \bigg\{  \hspace{-0.8mm}
\frac{ \sqrt{6} (436\hspace{-0.5mm}+\hspace{-0.5mm}93 \pi ) }{81(2\hspace{-0.5mm}+\hspace{-0.5mm}\pi )^{5/2}} \sin \xi
\hspace{-0.4mm} - \hspace{-0.4mm} \frac{2 \sqrt{6} }{27 (2\hspace{-0.5mm}+\hspace{-0.5mm}\pi )^{3/2}}[5 \cos 3 \xi  \hspace{-0.5mm}-\hspace{-0.5mm}  \sin 3 \xi] \hspace{-0.5mm} \bigg \} .  \nonumber 
\end{eqnarray}
The optimal value of the variational parameter for the angular frequency to first order is given by (\ref{K1}).  Inserting this value into (\ref{V1VPTOrd1}), (\ref{V2VPTOrd2}), we find the following VPT result for the limit cycle:
\begin{eqnarray}
V_{1,{\rm VPT}}^{(2)}(\xi) \hspace{-0.5mm} & \hspace{-2mm}=\hspace{-2mm}& \hspace{-0.5mm} \frac{1}{27 \sqrt{3 (2+\pi )} (2 + \pi + 4 g)^2} \big \{ 108 (2+\pi )^2 \cos \xi \nonumber \\ &&  \hspace{-18mm}+ \hspace{1mm} g \left[(1148+699 \pi )\cos \xi -6 (2+\pi ) (\cos 3 \xi -7 \sin 3 \xi)\right ] \big \}\, , \nonumber  \\ 
%\end{eqnarray}
\\
%\begin{eqnarray}
V_{2,{\rm VPT}}^{(2)}(\xi) \hspace{-0.5mm} & \hspace{-2mm} =  \hspace{-2mm} & \hspace{-0.5mm} \frac{2}{27 \sqrt{6
   (2+\pi )} (2 + \pi + 4 g)^2} \big \{ 108 (2+\pi )^2 \sin \xi  \nonumber \\ 
&& \hspace{-18mm} + \hspace{1mm} g \left[(1292+771 \pi ) \sin \xi +  6 (2+\pi )(5 \cos 3  \xi - \sin 3 \xi) \right] \big \} \, . \nonumber \\
\end{eqnarray}
The procedure in higher order is analogous. Figure \ref{fig6} {\bf b)} shows our VPT results for the limit cycle for $\epsilon = \nolinebreak[4] 2$ up to the eighth order. Figure \ref{fig7} {\bf b)} shows the logarithm of the error measure (\ref{ErMe}) for the VPT limit cycle versus the order $N$ for different values of $\epsilon$. In Fig.~\ref{fig8} {\bf b)} the accuracy of the eighth order results from the Shohat expansion and VPT are compared. Again, we find that our VPT results are more reliable than those from the Shohat expansion, especially for larger delays. \\
%%%%%%%%%%%%%%%%%%%%%%%%%%%%%%%%%%%%%%%%%%%%%
\section{Summary}
%%%%%%%%%%%%%%%%%%%%%%%%%%%%%%%%%%%%%%%%%%%%%
We have performed a perturbative calculation of the limit cycle and its frequency in a two-neuron model with delay. A Shohat resummation of the respective perturbation expansions yields results which are in good agreement with numerical values but whose accuracy decreases drastically with larger values of the delay parameter. Resumming the perturbation series with VPT yields more uniformly converging results, which are reliable even in low orders, and furthermore permits the extraction of the leading large-delay behavior with sufficient accuracy. 
The present work constitutes the first application of VPT to a system of DDE's.  Moreover, it establishes a method for the variational resummation of perturbatively calculated limit cycles in nonlinear dynamical systems.
%%%%%%%%%%%%%%%%%%%%%%%%%%%%%%%%%%%%%%%%%%%%%%%%%%%%%%%%%
\section{Acknowledgement}
%%%%%%%%%%%%%%%%%%%%%%%%%%%%%%%%%%%%%%%%%%%%%%%%%%%%%%%%%
We wish to acknowledge assistance from Michael Schanz in solving the system of DDE's (\ref{dde1}), (\ref{dde2}) numerically.  Simulations were carried out with the AnT 4.669 software \cite{AnT}.

We thank Kevin Archie, Carl Bender, John Clark,  Ulrich Kleinekath{\"o}fer, and Hagen Kleinert for critical reading of the manuscript.

This works was supported in part by NIH-EY 15678.

%%%%%%%%%%%%%%%%%%%%%%%%%%%%%%%%%%%%%%%%%%%%%%%%%%%%%%%%%
%%%%%%%%%%%%%%%%%%%%%%%%%%%%%%%%%%%%%%%%%%%%%%%%%%%%%%%%%
\begin{appendix}
%%%%%%%%%%%%%%%%%%%%%%%%%%%%%%%%%%%%%%%%%%%%%%%%%%%%%%%%%
\begin{widetext}
\section{Elimination of Secular Terms}
%%%%%%%%%%%%%%%%%%%%%%%%%%%%%%%%%%%%%%%%%%%%%%%%%%%%%%%%%
We now demonstrate how the conditions (\ref{Cond1}), (\ref{Cond2}) are obtained by considering the Fourier decompositions of the periodic solution and the inhomgeneity. Inserting (\ref{VFouExp}) and (\ref{fFouExp}) into the system of equations (\ref{V1nDE}), (\ref{V2nDE}) and comparing coefficients of $\sin k \xi$ and $\cos k \xi$ in both components, we obtain the following system of four equations:
\begin{eqnarray}
\frac{a_{1,k}^{(n)}}{\omega_0} + k b_{1,k}^{(n)} - \frac{a_1 a_{2,k}^{(n)}}{\omega_0} \cos (k \omega_0 \tau_0) + \frac{a_1 b_{2,k}^{(n)}}{\omega_0} 
\sin (k \omega_0 \tau_0) 
% \nonumber \\ 
- \alpha_{1,k}^{(n)} &=& 0 \, , 
%\hspace{4mm} 
\label{SecTerk1}\\
\frac{b_{1,k}^{(n)}}{\omega_0} - k a_{1,k}^{(n)} - \frac{a_1 b_{2,k}^{(n)}}{\omega_0} \cos (k \omega_0 \tau_0) - \frac{a_1 a_{2,k}^{(n)}}{\omega_0} 
\sin (k \omega_0 \tau_0) 
% \nonumber \\ 
- \beta_{1,k}^{(n)} &=& 0 \, ,  
% \hspace{4mm} 
\label{SecTerk2}
\\
\frac{a_{2,k}^{(n)}}{\omega_0} + k b_{2,k}^{(n)} - \frac{a_2 a_{1,k}^{(n)}}{\omega_0} \cos (k \omega_0 \tau_0) + \frac{a_2 b_{1,k}^{(n)}}{\omega_0} 
\sin (k \omega_0 \tau_0) 
% \nonumber \\ 
- \alpha_{2,k}^{(n)} &=& 0 \, ,  
% \hspace{4mm} 
\label{SecTerk3}
\\
\frac{b_{2,k}^{(n)}}{\omega_0} - k a_{2,k}^{(n)} - \frac{a_2 b_{1,k}^{(n)}}{\omega_0} \cos (k \omega_0 \tau_0) - \frac{a_2 a_{1,k}^{(n)}}{\omega_0} 
\sin (k \omega_0 \tau_0) 
% \nonumber \\ 
- \beta_{2,k}^{(n)} &=& 0 \, .  
% \hspace{4mm} 
\label{SecTerk4}
\end{eqnarray}
It turns out that for $k>1$ the coefficients $\mathbf{a}_k^{(n)}, \, \mathbf{b}_k^{(n)}$ can be uniquely determined for any inhomogeneity, i.e., for arbitrary $\boldsymbol{\alpha}_k^{(n)}, \, \boldsymbol{\beta}_k^{(n)}$.  For $k>1$ the solution of the system (\ref{SecTerk1}) -- (\ref{SecTerk4}) is
\begin{eqnarray}
a_{1,k}^{(n)} &=& \frac{1}{D} \Big\{ 
(\alpha_{1,k}^{(n)} - k \omega_0 \beta_{1, k}^{(n)}) (\omega_0 + k^2 \omega_0^3)  
- a_1 \omega_0^2 \sin(k \omega_0 \tau_0) (2 k \alpha_{2,k}^{(n)} - (1 + k^2) \omega_0 \beta_{2,k}^{(n)}) + a_1 \omega_0 \cos( k \omega_0 \tau_0)
\\
&& \hspace{-11mm} \times \hspace{0.8mm} (2 \alpha_{2,k}^{(n)} - 2 k \omega_0 \beta_{2,k}^{(n)}+ (1- k^2) \omega_0^2 \alpha_{2,k}^{(n)})
+(\omega_0 + \omega_0^3) \Big[
\sin(2 k \omega_0 \tau_0)(\beta_{1,k}^{(n)} - k \omega_0 \alpha_{1,k}^{(n)})
+\cos( 2 k \omega_0 \tau_0)(\alpha_{1,k}^{(n)} + k \omega_0 \beta_{1,k}^{(n)})
\Big]\Big\} \, , \nonumber
\\
b_{1,k}^{(n)} &=& \frac{1}{D} \Big\{ 
(\beta_{1,k}^{(n)} + k \omega_0 \alpha_{1, k}^{(n)}) (\omega_0 + k^2 \omega_0^3)
- a_1 \omega_0^2 \sin(k \omega_0 \tau_0) (2 k \beta_{2,k}^{(n)} + (1 + k^2) \omega_0 \alpha_{2,k}^{(n)}) + a_1 \omega_0 \cos( k \omega_0 \tau_0)
\\
&& \hspace{-11mm} \times \hspace{0.8mm} (2 \beta_{2,k}^{(n)} + 2 k \omega_0 \alpha_{2,k}^{(n)}+ (1- k^2) \omega_0^2 \beta_{2,k}^{(n)})
+(\omega_0 + \omega_0^3) \Big[
\cos(2 k \omega_0 \tau_0)(\beta_{1,k}^{(n)} - k \omega_0 \alpha_{1,k}^{(n)})
-\sin( 2 k \omega_0 \tau_0)(\alpha_{1,k}^{(n)} + k \omega_0 \beta_{1,k}^{(n)})
\Big]\Big\} \, , \nonumber
\end{eqnarray}
%\\
\begin{eqnarray}
a_{2,k}^{(n)} &=& \frac{1}{D} \Big\{ 
(\alpha_{2,k}^{(n)} - k \omega_0 \beta_{2, k}^{(n)}) (\omega_0 + k^2 \omega_0^3) + 
(\omega_0 + \omega_0^3) \Big[ 
\frac{\omega_0}{a_1} \sin (k \omega_0 \tau_0) (2 k \alpha_{1,k}^{(n)} - \omega_0(1 + k^2) \beta_{1,k}^{(n)}) 
\\ && \hspace{-10mm} - \cos(k \omega_0 \tau_0)( 2 \alpha_{1,k}^{(n)} - 2 k \omega_0 \beta_{1,k}^{(n)} + (1 - k^2)\omega_0^2 \alpha_{1,k}^{(n)})
+ \sin(2 k \omega_0 \tau_0)(\beta_{2,k}^{(n)} - k \omega_0 \alpha_{2,k}^{(n)})
+ \cos(2k \omega_0 \tau_0)(\alpha_{2,k}^{(n)} + k \omega_0 \beta_{2,k}^{(n)})
\Big]\Big\}\, , \nonumber
\\
b_{2,k}^{(n)} &=& \frac{1}{D} \Big\{ 
(\beta_{2,k}^{(n)} + k \omega_0 \alpha_{2, k}^{(n)}) (\omega_0 + k^2 \omega_0^3) + 
(\omega_0 + \omega_0^3) \Big[ 
\frac{\omega_0}{a_1} \sin (k \omega_0 \tau_0) (2 k \beta_{1,k}^{(n)} + \omega_0(1 + k^2) \alpha_{1,k}^{(n)}) 
\\ && \hspace{-10mm} - \cos(k \omega_0 \tau_0)( 2 \beta_{1,k}^{(n)} + 2 k \omega_0 \alpha_{1,k}^{(n)} + (1 - k^2)\omega_0^2 \beta_{1,k}^{(n)})
- \sin(2 k \omega_0 \tau_0)(\alpha_{2,k}^{(n)} + k \omega_0 \beta_{2,k}^{(n)})
+ \cos(2k \omega_0 \tau_0)(\beta_{2,k}^{(n)} - k \omega_0 \alpha_{2,k}^{(n)})
\Big]\Big\}\, , \nonumber
\end{eqnarray}
where
\begin{eqnarray}
D 
%&
=
%& 
2 + 2\omega_0^2(1 + k^2) + \omega_0^4(1 + k^4) + (\omega_0 + \omega_0^3)
%\\ 
%&& 
%\hspace{-0mm} \times \hspace{1mm} 
\big[ 2 (1 - k^2 \omega_0^2) \cos(2 \omega_0 \tau_0) / \omega_0 - 4k \sin(2 k \tau_0 \omega_0)\big] \, . 
%\nonumber
\end{eqnarray}
Note that $D$ vanishes for $k = 1$. 
We must thus reconsider the system (\ref{SecTerk1}) -- (\ref{SecTerk4}) for the case $k=1$ and it turns out that $\boldsymbol{\alpha}_1^{(n)}, \, \boldsymbol{\beta}_1^{(n)}$ must satisfy  certain conditions for a solution to exist.  For $k=1$, we add $a_2 \sin(\omega_0 \tau_0) \hspace{1mm} \times$ (\ref{SecTerk1}) to (\ref{SecTerk4}) and subtract $\omega_0 \hspace{1mm} \times$ (\ref{SecTerk3}) from $a_2 \cos(\omega_0 \tau_0) \hspace{1mm} \times$ (\ref{SecTerk2}). Using the identities $a_1 a_2 = - (\omega_0^2 + 1)$ and $\omega_0 = \cot(\omega_0 \tau_0)$, we obtain the two conditions (\ref{Cond1}), (\ref{Cond2}) that must be satisfied by the inhomgeneity ${\bf f}^{(n)}(\xi)$.
Imposing (\ref{Cond1}), (\ref{Cond2}) on ${\bf f}^{(n)}(\xi)$, we obtain the following solution to the system of equations (\ref{SecTerk1}) -- (\ref{SecTerk4}) for $k=1$:
\begin{eqnarray}
b_{1,1}^{(n)} &=& \cos(\omega_0 \tau_0)\left[a_2 \cos(\omega_0 \tau_0) \alpha_{1,1}^{(n)} + \frac{\alpha_{2,1}^{(n)}}{a_2}\right] 
%\nonumber \\ && 
%- \hspace{1mm} 
\frac{a_{2,1}^{(n)}}{a_2 \sin(\omega_0 \tau_0)} \, , 
\\
%\end{eqnarray}
%\begin{eqnarray}
b_{2,1}^{(n)} &=& a_2 \sin(\omega_0 \tau_0) \left[a_{1,1}^{(n)} - \alpha_{1,1}^{(n)} \sin(\omega_0 \tau_0) \cos(\omega_0 \tau_0)\right] 
%\nonumber \\&& 
+ 
%\hspace{1mm} 
\alpha_{2,1}^{(n)} \cos^2(\omega_0\tau_0) \, .
\end{eqnarray}
Here, the coefficients $a_{1,1}^{(n)}$, $a_{2,1}^{(n)}$ are undetermined and follow from the initial conditions.  We set $a_{1,1}^{(n)} = A_n$ and $a_{2,1}^{(n)} = 0$.
\end{widetext}
\end{appendix}

%%%%%%%%%%%%%%%%%%%%%%%%%%%%%%%%%%%%%%%%%%%%%%%%%%%%%%%%%

\end{document}